\documentclass[10pt,twocolumn,letterpaper]{article}

\usepackage[pagenumbers]{cvpr} 

\usepackage{tocloft}
\usepackage{booktabs}
\usepackage{multirow}
\usepackage[table]{xcolor}
\usepackage{colortbl}

\advance\cftsecnumwidth 0.5em\relax
\advance\cftsubsecindent 0.5em\relax
\advance\cftsubsecnumwidth 0.5em\relax

\usepackage{mathtools}
\usepackage{titletoc}
\usepackage[ruled,vlined]{algorithm2e}
\SetAlFnt{\small}                
\SetAlCapFnt{\small}             
\SetAlCapNameFnt{\small}
\SetAlgoInsideSkip{0pt}          
\SetAlCapSkip{0.3em}             
\SetKwInput{KwInput}{Input}
\SetKwInput{KwOutput}{Output}

\usepackage{cancel}
\usepackage{bbm}
\newcommand{\ind}[1]{\mathbbm{1}_{\{#1\}}}

\usepackage[dvipsnames]{xcolor}
\usepackage[table]{xcolor}

\usepackage[most]{tcolorbox}
\usepackage{listings}
\usepackage{multirow}

\newcommand{\ignore}[1]{}
\usepackage{booktabs}

\definecolor{codegreen}{rgb}{0,0.6,0}
\definecolor{codegray}{rgb}{0.5,0.5,0.5}
\definecolor{codepurple}{rgb}{0.58,0,0.82}
\definecolor{backcolour}{rgb}{0.95,0.95,0.92}
\definecolor{darkgreen}{rgb}{0,0.5,0}

\lstdefinestyle{mystyle}{
    backgroundcolor=\color{backcolour},   
    commentstyle=\color{codegreen},
    keywordstyle=\color{codepurple},
    numberstyle=\tiny\color{codegray},
    stringstyle=\color{codepurple},
    basicstyle=\ttfamily\footnotesize,
    breakatwhitespace=false,         
    breaklines=true,                 
    captionpos=b,                    
    keepspaces=true,                                  
    numbersep=5pt,      
    frame = shadowbox,
    showspaces=false,                
    showstringspaces=false,
    showtabs=false,                  
    tabsize=2
}

\newcommand{\boxedthm}[1]{
\begin{tcolorbox}[breakable, enhanced,
                  colback=gray!30,
                  colframe=black,
                  width=\linewidth,
                  arc=2mm, auto outer arc,
                  boxrule=1pt,
                  boxsep=-1mm,
                 ]
  #1
\end{tcolorbox}
}

\newcommand{\boxedeg}[1]{
\begin{tcolorbox}[colback=orange!10,
                  colframe=black,
                  width=\linewidth,
                  arc=2mm, auto outer arc,
                  boxrule=1pt,
                  boxsep=-1mm,
                 ]
  #1
\end{tcolorbox}
} 

\usepackage{amsthm}

\newtheoremstyle{withdot}%
  {\topsep}{\topsep}
  {\itshape}
  {}
  {\bfseries}
  {.}
  {.5em}
  {\thmname{#1}\thmnumber{ #2}\thmnote{ \bfseries(#3)}}

\theoremstyle{withdot}

\newtheorem{theorem}{Theorem}
\newtheorem{proposition}{Proposition}
\newtheorem{lemma}{Lemma}

\definecolor{cvprblue}{rgb}{0.21,0.49,0.74}
\usepackage[pagebackref,breaklinks,colorlinks,allcolors=cvprblue]{hyperref}

\def\paperID{***} 
\def\confName{***}
\def\confYear{***}

\title{Markov-Renewal Single-Photon LiDAR Simulator}

\author{Weijian Zhang
\quad
Prateek Chennuri
\quad
Hashan K. Weerasooriya
\quad
Bole Ma
\quad
Stanley H. Chan\\
School of Electrical and Computer Engineering, Purdue University\\
{\tt\small \{zhan5056, pchennur, hweeraso, ma929,  stanchan\}@purdue.edu}
}

\begin{document}

\twocolumn[{%
\renewcommand\twocolumn[1][]{#1}%
\maketitle
\begin{center}
    \vspace{-2.5em}
    \captionsetup{type=figure}
        \includegraphics[width=0.95\textwidth]{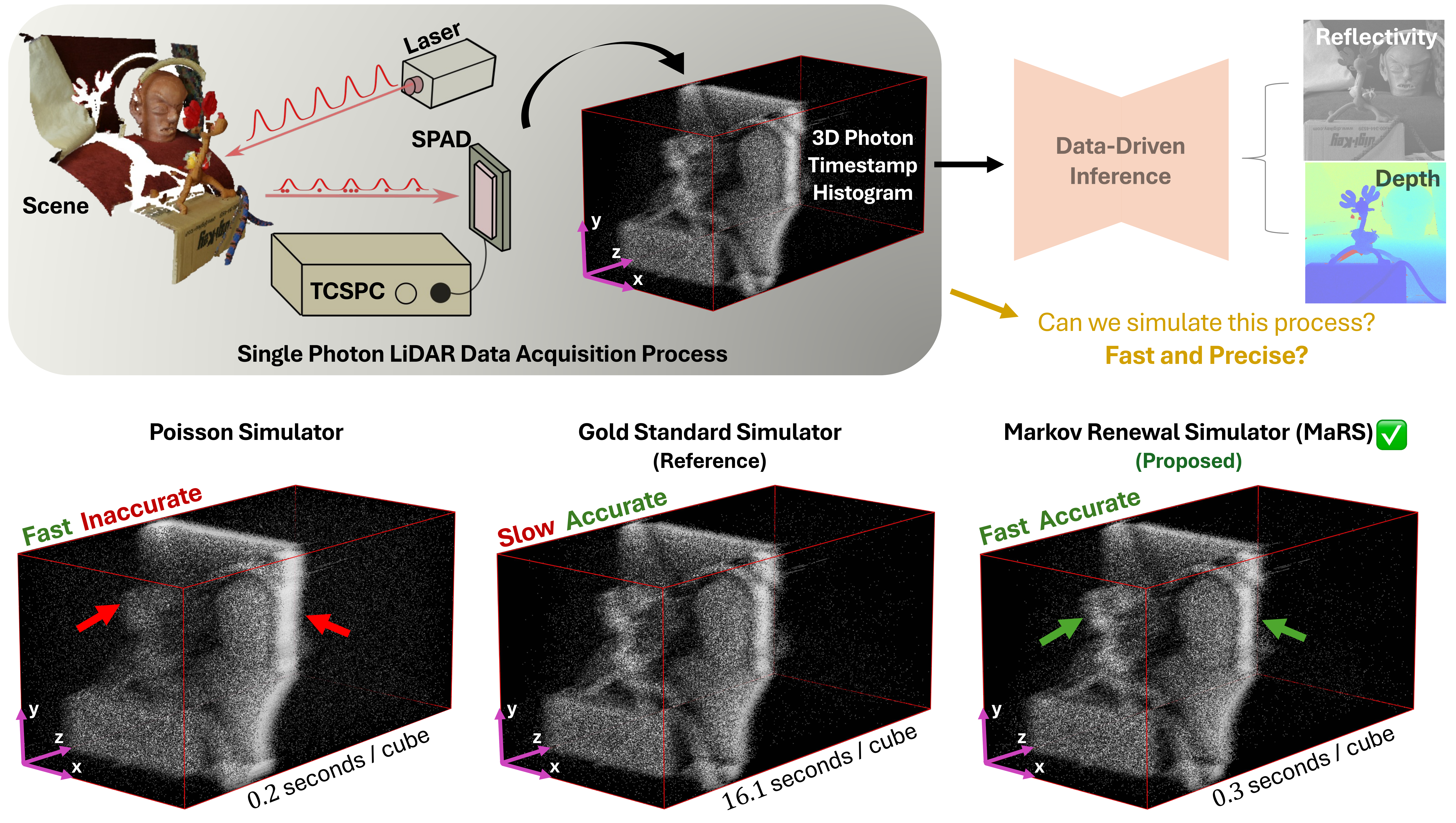}
    \\
    \vspace{-0.5em}
    \captionof{figure}{
    \textbf{The SP-LiDAR Simulation Dilemma: Speed vs. Accuracy.}
    The physical acquisition process (top) forces a trade-off. 
    \textbf{(Left)} The Poisson model is fast (0.2s) but physically wrong, as it ignores detector dead time. 
    \textbf{(Center)} The sequential ``gold-standard" is physically-correct, modeling dead time, but is computationally intractable (16.1s). 
    \textbf{(Right)} Our Markov-renewal simulator analytically models dead time, achieving the same high fidelity as the gold-standard at a practical compute cost (0.3s). Our framework breaks the compromise, enabling fast and physically correct simulation. 
    }
    \label{fig:teaser}
\end{center}
}]

\maketitle
\begin{abstract}
Single-photon LiDAR (SP-LiDAR) simulators face a dilemma: fast but inaccurate Poisson models or accurate but prohibitively slow sequential models. This paper breaks that compromise. We present a simulator that achieves both fidelity and speed by focusing on the critical, yet overlooked, component of simulation: the photon count statistics. Our key contribution is a Markov-renewal process (MRP) formulation that, for the first time, analytically predicts the mean and variance of registered photon counts under dead time. To make this MRP model computationally tractable, we introduce a spectral truncation rule that efficiently computes the complex covariance statistics. By proving the shift-invariance of the process, we extend this per-pixel model to full histogram cube generation via a precomputed lookup table. Our method generates 3D cubes indistinguishable from the sequential gold-standard, yet is orders of magnitude faster. This finally enables large-scale, physically-faithful data generation for learning-based SP-LiDAR reconstruction.
\end{abstract}    
\section{Introduction}
\label{sec:intro}

\textbf{Motivation.}
Single-photon LiDAR (SP-LiDAR) is a key technology for high-resolution imaging \cite{Gyongy_2022_dtof, Piron_2021_review_spad, Morimoto_2020_SPAD, Xuanyu_2024_spad_automotive_challenges, Li_2020_LiDARReview, young_2025_enhancingautonomousnavigationimaging, Li_2023_SPAD, Degnan_2016_bathymetric, Li_2021_200km_imaging, McCarthy_2013_kmrange, McCarthy_2025_long_distance, under_water_2023_Maccarone, halimi_2017_joint_underwater, Jiang_23_long_obscurants, fog_2022_zhang, qian_2023_modelling,kavinga_2025_joint_estimation}.
Its single-photon sensitivity, however, introduces complex measurement statistics. Photon detections follow \emph{stochastic point-process dynamics} \cite{qian_2023_modelling} that are fundamentally coupled to detector state.
Specifically, single-photon avalanche diodes (SPADs) enter \emph{dead time} after each registration, ignoring subsequent photons. This effect is non-linear: it destroys the statistical independence of detections and creates complex temporal dependencies.
The result is a fundamental distortion of both the photon count and the temporal probability density function (PDF), invalidating simple Poisson models~\cite{zhang_2025_icip}. Any simulator that ignores this is physically-wrong.
\vspace{0.5em}

\noindent \textbf{Limitations of existing simulators.}
The field is paralyzed by a false dichotomy.
On one hand, Poisson-multinomial models are fast but physically-incorrect, as they ignore dead time.
On the other hand, sequential ``gold-standard" simulators are physically-correct, as they brute-force the rejection sampling of photons during dead time.
This sequential approach is computationally intractable, requiring per-photon loops that do not scale in flux, resolution, or realizations.
It is a bottleneck that has prevented the generation of large-scale, physically-correct datasets required for training modern deep learning pipelines.
\vspace{0.5em}

\noindent \textbf{Contribution.}
This paper resolves the speed-vs-accuracy conflict. We present the \textbf{Markov Renewal SP-LiDAR Simulator (MaRS)}.
The core idea is to \emph{decouple} the simulation problem: the temporal PDF is well-studied, but the primary bottleneck is the \emph{photon count}, which prior work has failed to model efficiently.
We theoretically show that by modeling the count statistics as a \emph{Markov-renewal process} (MRP)~\cite{cinlar_1969_mrp,Fox1968SEMIMARKOV}, we can preserve tractability while rigorously accounting for the state-dependent correlations caused by dead time.
This MRP formulation is the key, providing an analytic path to the count statistics that bypasses the intractable sequential simulation.
Our contributions are two fold:
\begin{itemize}
    \item A new theoretical model of the photon count statistics under dead time using a Markov-renewal process (MRP).
    \item A new, high-throughput simulator (MaRS) that implements this model using spectral truncation and pixel parallelization.
\end{itemize}

\noindent As shown in Figure~\ref{fig:teaser}, MaRS produces physically-correct histogram cubes at a computation cost on par with the physically-incorrect Poisson model, and orders of magnitude faster than the gold-standard.
It handles all flux regimes and is the first SP-LiDAR simulator that provides the accuracy of a sequential model at the speed of an analytic one.

\section{Related Work}
\label{sec:related_work}

\textbf{Existing Simulators.} The \emph{gold standard} or \emph{conventional simulator}, first generates photon arrivals using multinomial Poisson sampling, and then sequentially rejects detections falling within the dead-time window~\cite{Rapp_2019_Dead,zhang_2024_mmsp}.
While physically accurate, this sequential rejection requires per-photon, per-pixel iteration and is therefore computationally prohibitive for megapixel arrays or large-scale data generation.

Zhang \etal\ recently attempted to address this bottleneck by directly modeling dead-time-distorted statistics, incorporating both the temporal PDF and photon count~\cite{zhang_2025_icip}. 
However, their autoencoder-based estimation of the temporal PDF must be retrained for each dead-time setting, their count prediction exhibits limited accuracy in capturing variance, and the simulation pipeline remains pixel-wise rather than fully parallel. 
In contrast, our simulator provides a unified and flexible framework that enables fully parallel synthesis of entire histogram cubes with improved statistical fidelity across different operating conditions.

\noindent \textbf{Statistical Modeling in Presence of Dead Time.}
Faithful histogram simulation demands two things: the dead-time-distorted temporal PDF and the total photon count.

\underline{Temporal PDF.}
Modeling synchronous systems is a solved problem; their independent increments are trivial to model~\cite{heide_2018_sub,Pediredla_2018_pileup,Gupta_2019_Asynchronous,Pediredla_2022_adaptive_gating}. The real challenge is \emph{asynchronous} operation~\cite{Kao_2025_freerunning,lyu_2025_highcounting}, where stochastic arrivals and dead time create complex, dependent increments.
Prior work addressed this using Markov chains to model relative registration states~\cite{Rapp_2019_Dead,rapp_2021_high}, later accelerated in~\cite{zhang_2025_PDFacceleration}.
This line of work is useful, but it \emph{only} predicts the temporal PDF. It offers no insight into \emph{how many} photons are registered, making it an incomplete solution for a full simulator.

\underline{Photon count.}
This remains an open research problem. Modeling the count $N(t)$ requires tracking detections in \emph{absolute} time.
The aforementioned Markov chain framework is the wrong tool for this job; it captures relative transitions, discarding the absolute timing needed to count events.
Other frameworks are no better: self-exciting point processes are computationally intractable~\cite{Snyder_1991_book}, and standard renewal theory~\cite{Yu_2000_counting,Gupta_2019_Asynchronous} breaks down because our inter-registration times ${W_k}$ are dependent, not independent and identically distributed (i.i.d).
We bridge this gap with a \emph{Markov–renewal process} (MRP).
The MRP formulation is the first to preserve renewal-like tractability while correctly handling the state-dependent correlations. This is the machinery required to derive the statistics of the registration times ${T_n}$ and, by inversion, solve for the photon count $N(t)$.

\textbf{Utility for Downstream Tasks.} Simulators are typically validated through downstream reconstruction tasks.
While depth recovery has been widely explored~\cite{Lindell_2018_SIGGRAPH,peng_2020_non_local, sun_2020_monocular_fusion, Tachella_2019_pnp_3d_reconstruction, altmann_2020_individual}, reflectivity estimation from raw photon histograms remains far less studied despite its importance for material inference and albedo calibration~\cite{vivek_2024_detection}.
Accurate count modeling is essential for reflectivity recovery, as the expected number of photon registrations directly encodes surface brightness.
By producing physically consistent photon counts under dead time, our simulator not only enables reliable reflectivity estimation but also establishes a challenging new benchmark for evaluating such algorithms.

\section{MaRS: Design Philosophy}
\label{sec:bg}
The goal of MaRS is to provide a unified simulation framework that remains both physically accurate and computationally efficient.
Rather than simulating every photon arrival individually, MaRS formulates the LiDAR measurement process through analytical and statistical models that explicitly account for detector dead time and temporal correlations.

\begin{figure*}[t]
    \centering
    \includegraphics[width=\linewidth]{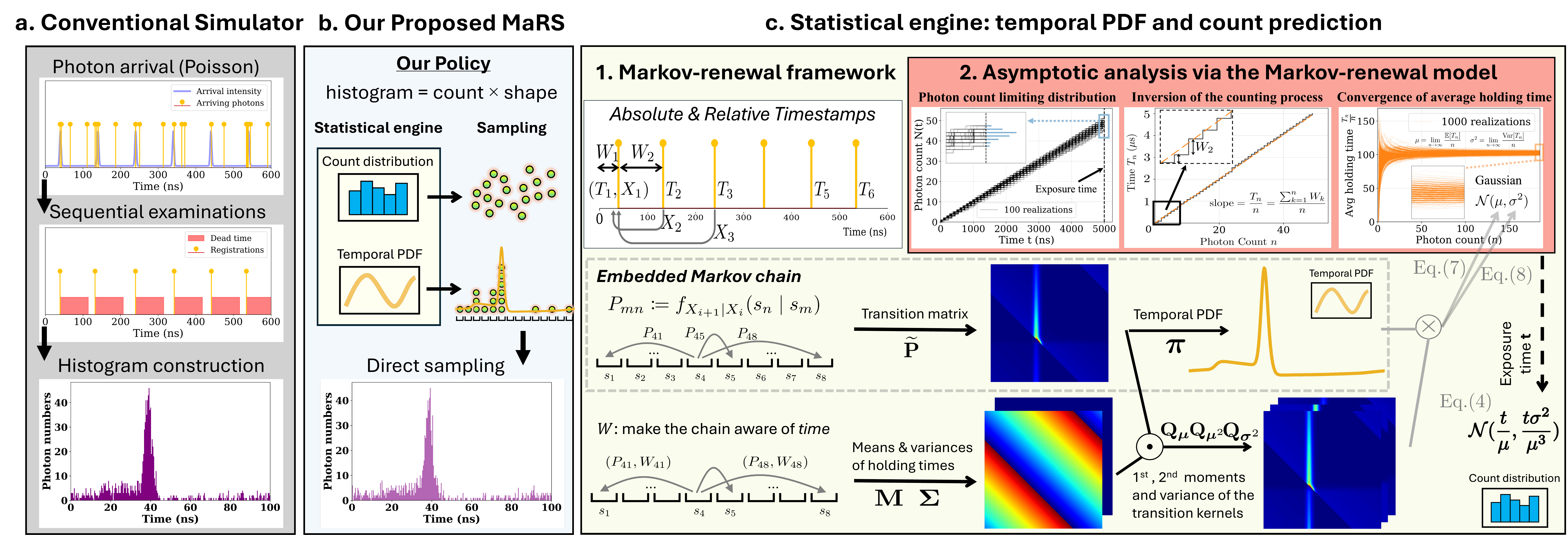}
    \caption{
    \textbf{Proposed MaRS Simulator Overview.}
    (a) Traditional high-flux LiDAR simulation requires sequential Poisson sampling, dead-time enforcement, and timestamp binning—physically accurate but slow and hard to scale. 
    (b) MaRS models a histogram as a product of \emph{count} and \emph{shape}. A statistical engine predicts both the photon-count distribution and temporal PDF, enabling \emph{one-shot sampling} without step-wise time simulation. 
    (c) The core engine is built on a \emph{Markov–renewal process}: 
    (c1) an embedded Markov chain with state-dependent holding times captures temporal correlations and yields the stationary temporal PDF; 
    (c2) asymptotic analysis shows these holding times converge to a Gaussian with closed-form mean and variance, producing a Gaussian photon-count law. 
    Together, these enable fast, analytic, and accurate prediction of both histogram \emph{scale} and \emph{shape}.
    }
    \label{fig:overview_mrp}
\end{figure*}

\subsection{Dead-Time-Distorted Forward Model}
\label{subsec:notation}
The measurement process in single-photon LiDAR can be viewed as two sequential stages: photons first \emph{arrive} at the detector and are then \emph{registered} by the sensor, forming a histogram over one laser repetition period.

\noindent \textbf{Photon arrivals.}
Following prior work \cite{Chan_2024_CVPR, rapp_2017_unmixing, Bar-David_1969, Shin_2015_3D}, we model photon arrivals as an inhomogeneous Poisson process with a per-cycle rate
\begin{equation}
    \label{eq:arrival_flux}
    \lambda(t) = \alpha\, s(t-\tau) + \lambda_b,
    \quad 0 \leq t < t_r,
\end{equation}
where $s(t)$ is the emitted laser pulse, $\alpha$ is the surface reflectivity, $\tau$ is the depth-induced delay, and $\lambda_b$ is the constant ambient light. The transmitted pulse is modeled as a Gaussian $s(t) = \calN(t;0,\sigma_t^2)$. We assume that $N_r$ pulses are emitted, each separated by a repetition period $t_r$, resulting in a total exposure time of $N_rt_r$. The signal level is defined as $S = \alpha$, the noise level as $B = t_r \lambda_b$, and the total flux as $\Lambda = S+B$. The signal-to-background ratio (SBR) is given by $\text{SBR} = S/B$.

The arrival flux governs the ideal histogram when detector
nonlinearities are absent, as both the \emph{number} of photons and the \emph{temporal
shape} scale linearly with the flux.

\noindent \textbf{Registration under dead time.}
A SPAD introduces a dead time $t_d$ after each detection, during which incoming photons are lost.  
In asynchronous operation, the detector becomes active again immediately after $t_d$.
The TCSPC module records timestamps $X_k = T_k \bmod t_r$ to one of its $n_b$ bins, where $T_k$ are the absolute detection times.  
When operating under dead time, the recorded timestamps deviate from the true arrival distribution: temporal correlations arise, the photon count rate decreases, and the histogram becomes distorted.

\noindent \textbf{Input parameters.}
We define the complete system configuration as
$\vtheta=(t_r, N_r, \sigma_t, t_d, n_b, \tau, S, B, N_\mathrm{iter})$, where $N_\mathrm{iter}$ represents the number of histogram realizations to be generated. In the presence of dead time, the histogram becomes a nonlinear function of $\vtheta$.
Since closed-form models for the joint distribution of all histogram bins are typically unavailable under free-running, inhomogeneous conditions, a decoupled modeling strategy is addopted in the following section.

\subsection{Histogram Synthesis Decoupling}
\label{subsec:temporal_PDF}
Unlike the conventional two-step simulation, which is often computationally inefficient, we decouple the histogram synthesis process into two components: photon count and temporal shape, as illustrated in \cref{fig:overview_mrp}. Specifically, we first estimate the distorted photon count and the corresponding temporal probability density function (PDF), and then directly sample histograms based on these statistics.

The conventional simulator iterates through all photons and realizations sequentially to produce different histogram realizations for the same scene. In contrast, the proposed simulation framework relies on statistical modeling. Once the model is established, it can be reused to generate multiple histograms in parallel, as summarized in \cref{alg:single_pixel_sim}.

\begin{algorithm}[t]
\caption{Single-Pixel Histogram Generation}
\label{alg:single_pixel_sim}
\small
\DontPrintSemicolon
\KwIn{$\vtheta=(t_r, N_r, \sigma_t, t_d, n_b, \tau, S, B, N_\mathrm{iter})$.}
\KwOut{$N_\mathrm{iter}$ histograms $\vh_i \in \mathbb{R}^{N_\mathrm{iter} \times n_b}$.}

\textbf{Step 1: Statistical model.}  
Find the photon count distribution $p_N(n; \vtheta)$ and the temporal PDF $\vpi(\vtheta)$.\;

\textbf{Step 2: Photon count sampling.}  
Draw $N_\mathrm{iter}$ i.i.d. photon counts:  
$n_i \sim p_N(n; \vtheta)$.\;

\textbf{Step 3: Histogram synthesis.}  
Sample histograms via $\vh_i \sim \mathrm{Multinomial}(n_i, \vpi)$ for all realizations in parallel.\;
\end{algorithm}

\noindent \textbf{Stationary Temporal PDF.}
The Markov chain model proposed by Rapp \etal has been shown to accurately describe the stationary temporal PDF of photon registrations~\cite{Rapp_2019_Dead,rapp_2021_high}. It is a discrete approximation that divides the continuous repetition period $[0, t_r)$ into a set of states $\mathcal{S} = \{s_1,\dots,s_{n_b}\}$, where $n_b$ is the number of histogram bins. At its core is a transition matrix $\mPtilde$ that models the probability of transitions between the current and next states. Recently, Zhang \etal accelerated the reconstruction of a large transition matrix by a parallel reformulation and a separation of the role of dead time's in row shifting~\cite{zhang_2025_PDFacceleration}. As a result, the temporal PDF can be efficiently obtained by solving the leading left eigenvector $\vpi \mPtilde = \vpi$.

\noindent \textbf{Photon count: an open problem.}
Although the solution to the temporal PDF is available, the photon count modeling is still an open problem. Without accurately estimating it, the scale of the generated histogram fail to reflect the true statistics, introducing bias in sensor characterization or downstream neural network training. We solve this problem in \cref{sec:mrp}.
\section{Markov-Renewal Photon Count Prediction}
\label{sec:mrp}

The stationary temporal PDF describes \emph{where} timestamps are across histogram bins, but it does \emph{not} specify \emph{how many} photons are registered. 
The total photon count $N(t)$ is a random variable driven by inhomogeneous arrivals and detector dead time.
Classical Poisson or renewal models break down because inter-registration times are state-dependent and temporally correlated.

In this section, we develop a probabilistic model for the number of photon registrations.
We first identify the limitations of the Markov chain formulation and bridge the gap with Markov-renewal theory.
Then, we show that the photon count asymptotically converges to a Gaussian distribution.
Finally, we derive closed-form expressions for its mean and variance. Our method is summarized in~\cref{fig:overview_mrp}.

\subsection{From Markov Chain to Markov Renewal}
The photon registration process is a non-decreasing counting process $\{N(t): t \geq 0\}$ where $N(t)$ represents the number of photon registrations up to time $t$.
Let $T_k$ denote the \emph{absolute} registration timestamp of the $k$th photon, and $W_k$ the inter-registration time between the $(k-1)$th and the $k$th photons: $W_k = T_k - T_{k-1}$ for $k \geq 2$, and $W_1 = T_1$.
The \emph{relative} phase of the $k$th registration within one laser period $t_r$ is $X_k=T_k\bmod t_r$, ignoring quantization alignment to bin centers.

The Markov chain model in \cref{subsec:temporal_PDF} captures transitions among the \emph{relative} states ${X_k}$ and therefore accurately characterizes the stationary temporal PDF.
However, it discards the \emph{absolute} timing information contained in ${T_k}$ and thus cannot predict the total number of events during a given exposure.
While it can reproduce histogram \emph{shapes}, it cannot recover the correct histogram \emph{scale} determined by the underlying photon flux and system parameters.

To overcome this limitation, we formulate the registration process as a \emph{Markov–renewal process (MRP)}: a stochastic process $\{(X_k, W_k)\}_{k \in \mathbb{N}}$ that jointly models the Markovian state transitions and the random holding times between them on the absolute scale.

The transition kernel of the MRP is a joint probability that the process jumps to state $j$ within time $t$, conditioned on state $i$:
\begin{align}
    Q_{ij}(t)
    & = \Pr(X_{k+1} = j, W_{k+1} \leq t \mid X_k = i) \label{eq:transition_kernel} \\
    & \bydef \underset{\Ptilde_{ij}}{\underbrace{\Pr(j \mid i)}} \, \underset{\coloneq F_{ij}(t)}{\underbrace{\Pr(W_{k+1} \leq t \mid i \rightarrow j)}},
    \label{eq:transition_kernel_bayes}
\end{align}
where $\Ptilde_{ij}$ is the transition probability mentioned in Sec.~\ref{subsec:temporal_PDF}, and $F_{ij}(t)$ is the conditional cumulative density function (CDF) of the holding time for transition $i \rightarrow j$.

This formulation allows the application of central limit theorem (CTL) arguments to the $n$th registration $T_n=\sum_{k=1}^n W_k$ and, by \emph{inversion}, to the photon count $N(t)$.


\begin{theorem}[Convergence of photon count $N(t)$]
    \label{thm:CLT_Nt}
    The photon count $N(t)$ asymptotically converges in distribution to a Gaussian:
    \begin{equation}
        \label{eq:CLT_Nt}
        N(t) \xrightarrow{d} \mathcal{N}\left(\frac{t}{\mu},\frac{t \sigma^2}{\mu^3}\right), \qquad t\to\infty,
    \end{equation}
    where the effective mean and variance of the holding times $\{W_k\}_{k \in \mathbb{N}}$ are defined as
    \begin{align}
        \mu
        & \bydef \lim_{n \rightarrow \infty} \frac{\E[T_n]}{n} = \lim_{n \rightarrow \infty} \frac{\E\left[\, \sum_{k=1}^n W_k \, \right]}{n}, \label{eq:equi_mean_def} \\
        \sigma^2
        & \bydef \lim_{n \rightarrow \infty} \frac{\Var[T_n]}{n} = \lim_{n \rightarrow \infty} \frac{\Var\left[\, \sum_{k=1}^n W_k \, \right]}{n}. \label{eq:equi_var_def}
    \end{align}
    (Proof in Supplement; obtained from the CLT for $T_n$ and inversion.)
\end{theorem}

\noindent\textbf{Remark.} \emph{
Our MRP model unifies several classical cases. For homogeneous Poisson arrivals without dead time, $W_k$ are i.i.d.\ exponential, $T_n$ is Erlang-distributed, and $N(t)$ is Poisson. With constant flux but fixed dead time, \cref{thm:CLT_Nt,thm:effective_mean,thm:effective_var} reduce to the standard delayed renewal results. Detailed derivations are provided in the Supplement.
}

\subsection{Closed-Form Equations for Mean and Variance}
\label{subsec:mu-sigma}

Although \cref{thm:CLT_Nt} establishes the asymptotic law for the photon count $N(t)$, we still require explicit expressions for $t$, $\mu$, and $\sigma^2$ in terms of the input parameters $\vtheta$.

The system run time is given by the total exposure time $t = N_r t_r$. Denote by $\vpi$ (row vector) the stationary distribution and by $\mPtilde$ the embedded transition matrix (both from Sec.~\ref{subsec:temporal_PDF}). To compute $\mu$ and $\sigma^2$, we construct from \cref{eq:transition_kernel_bayes} the first- and second-moment kernels $\mQ_{\vmu}\!=\mPtilde \odot \mM$, $\mQ_{\vmu^2}\!=\mPtilde \odot \mM \odot \mM$, and $\mQ_{\vsigma^2}\!=\mPtilde \odot \mSigma$, where $\odot$ denotes element-wise multiplication and $\mM=(\mu_{ij}), \mSigma=(\sigma^2_{ij})$ contain the transition-dependent mean and variance of the holding times:
\[
    \mu_{ij} = \E[\, W \mid i \rightarrow j \, ], \quad
    \sigma^2_{ij} = \Var[\, W \mid i \rightarrow j \, ].
\]

\begin{lemma}
    \label{lemma:cond_mean_var_holding_time}
    Conditioned on transition $i \rightarrow j$, the mean and variance of the holding time $W$ are
    \[
        \mu_{ij} = t_d + t_{i^{\prime}j} + \frac{e^{-\Lambda}}{1-e^{-\Lambda}} t_r, \quad
        \sigma_{ij}^2 = \frac{e^{-\Lambda}}{(1-e^{-\Lambda})^2} t_r^2,
    \]
    where $\Lambda$ is the photon arrival flux in one period and $t_{i^{\prime}j} = (x_j - x_{i^\prime}) \bmod t_r$ is the minimum awake interval between the reactivation time $x_{i^\prime}$, and the next registration $x_j$.
    
    \noindent (Proof in Supplement; holding times combine a deterministic offset with a stochastic number of full repetition periods.)
\end{lemma}

Lemma~\ref{lemma:cond_mean_var_holding_time} provides all quantities needed to compute $\mM$ and $\mSigma$, and thus $\mQ_{\vmu}$, $\mQ_{\vmu^2}$, and $\mQ_{\vsigma^2}$, enabling direct evaluation of $\mu$ and $\sigma^2$.

\begin{theorem}[Effective mean]
    \label{thm:effective_mean}
    The effective mean holding time defined in \cref{eq:equi_mean_def} is
    \begin{equation}
        \label{eq:equi_mean}
        \mu = \vpi \mQ_{\vmu} \mathbf{1}.
    \end{equation}
    (Proof in Supplement; by the law of total expectation.)
\end{theorem}

\begin{theorem}[Effective variance]
    \label{thm:effective_var}
    The effective variance of holding times defined in \cref{eq:equi_var_def} is
    \begin{equation}
        \label{eq:equi_var}
        \sigma^2 = \sigss^2 + 2 \gammass,
    \end{equation}
    which includes steady-state variance and covariance terms given by
    \begin{align}
        \sigss^2
        & = \vpi (\mQ_{\vmu^2} + \mQ_{\vsigma^2}) \mathbf{1}  - \mu^2, \label{eq:var_ss} \\
        \gammass
        & = \textstyle\sum_{l=1}^{\infty} \gamma_l \approx \textstyle\sum_{l=1}^c \big(\vpi \mathbf{Q_{\vmu}} \mPtilde^{\, l-1} \mathbf{Q_{\vmu}} \mathbf{1} - \mu^2\big), \label{eq:covar_ss}
    \end{align}
    respectively, where $c$ denotes a cutoff such that $\gamma_l \approx 0$ for $l > c$. Intuition: As the order $l$ increases, the covariance term $\gamma_l\!\to0$ geometrically with a rate set by the spectral gap of $\mPtilde$ (see Lemma~\textcolor{cvprblue}{2} in Supplement).
    
    \noindent (Proof in Supplement; via variance expansion + mixing.)
\end{theorem}

\section{Accurate and Efficient Histogram Cube Synthesis}
\label{sec:pillars}

To generate histogram cubes accurately and at scale, two challenges must be addressed: modeling the long-lag statistical effects introduced by dead time, and avoiding pixel-by-pixel sequential generation for an image. This section introduces the mechanisms that resolve these issues—spectral modeling of the covariance structure for accuracy, and a delay-invariance property for efficient reuse of per-pixel statistics—enabling fast and faithful cube synthesis.

\subsection{Spectral Analysis of Long-Lag Covariances}

While closed-form expressions for the variance in \cref{eq:equi_var,eq:covar_ss} require summing the covariance series 
$\gamma_{\mathrm{ss}} = \sum_{l\ge 1}\gamma_l$, the terms $\{\gamma_l\}$ exhibit widely varying decay patterns across different $(S,B,t_d)$ settings. 
This makes deterministic truncation (choosing a universal cutoff $c$) unreliable: large $c$ is computationally expensive, while small $c$ introduces bias. 
A data-adaptive rule is thus required.

\vspace{0.5em}
\noindent\textbf{Spectral motivation.}
Zhang~\etal~\cite{zhang_2025_PDFacceleration} showed that the long-term dynamics of the transition matrix $\widetilde{\mP}$ are governed by the magnitude and phase of a few dominant eigenvalues.  
Thus, although the early covariance terms depend on \emph{all} eigenmodes, the long-range correlations are effectively controlled by only the leading ones.  
We exploit this insight to derive a compact spectral approximation for $\gamma_{\mathrm{ss}}$.

\vspace{0.5em}
\noindent\textbf{Low-rank spectral model.}
Let the leading $p$ eigenpairs of $\widetilde{\mP}$ be 
$\{(\lambda_i,\mathbf{v}_i,\mathbf{u}_i)\}_{i=1}^p$ 
with biorthogonality $\mU_p^\top \mV_p = \mathbf{I}_p$.  
We approximate $\widetilde{\mP}^{\,l-1} \approx \mV_p \mLambda_p^{\,l-1} \mU_p^\top$, which is accurate for large $l$ because the remaining eigenvalues mix rapidly toward~$0$.  
Define $\alpha_i \coloneqq (\vpi\,\mQ_{\vmu}\,\mV_p)_i$ and $\beta_i \coloneqq (\mU_p^\top \mQ_{\vmu}\mathbf{1})_i$.

\begin{proposition}[Spectral truncation rule]
\label{prop:spectral_truncation_main}
For a chosen exact cutoff $L$ and spectral rank $p$, the covariance series
$\gamma_{\mathrm{ss}}$ can be approximated as
\begin{equation}
\widehat{\gamma}_{\mathrm{ss}}(L,p)
=
\sum_{l=1}^{L}\gamma_l
+
\sum_{i=2}^{p}
\frac{\alpha_i \beta_i\, \lambda_i^{\,L}}{1-\lambda_i}.
\label{eq:spectral_truncation_rule}
\end{equation}
The first term accounts for short-range correlations (computed exactly), while the second provides a closed-form estimate of the asymptotic tail using dominant eigenmodes.
\end{proposition}

\noindent\textbf{Remark.} \emph{
See proof of \cref{eq:spectral_truncation_rule} in the Supplement. 
Early covariance terms encode complex local behavior and are inexpensive to compute directly.  
Long-range correlations, however, are dictated by the slowest eigenmodes of $\widetilde{\mP}$.  
The spectral tail in \cref{eq:spectral_truncation_rule} captures this effect analytically.}

\vspace{0.5em}
\noindent\textbf{Choice of $(L,p)$.}
Empirically, across a range of $(S,B,t_d)$ settings, eigenvalues beyond the fifth have magnitude $\lesssim 0.7$, making their contribution negligible after $L\!\approx\!6$ steps.  
We therefore set $L=6$ and retain $p=5$ modes, which balances accuracy and runtime for all tested regimes. See more details in the Supplement.

\begin{figure}[t]
    \centering
    \includegraphics[width=\linewidth]{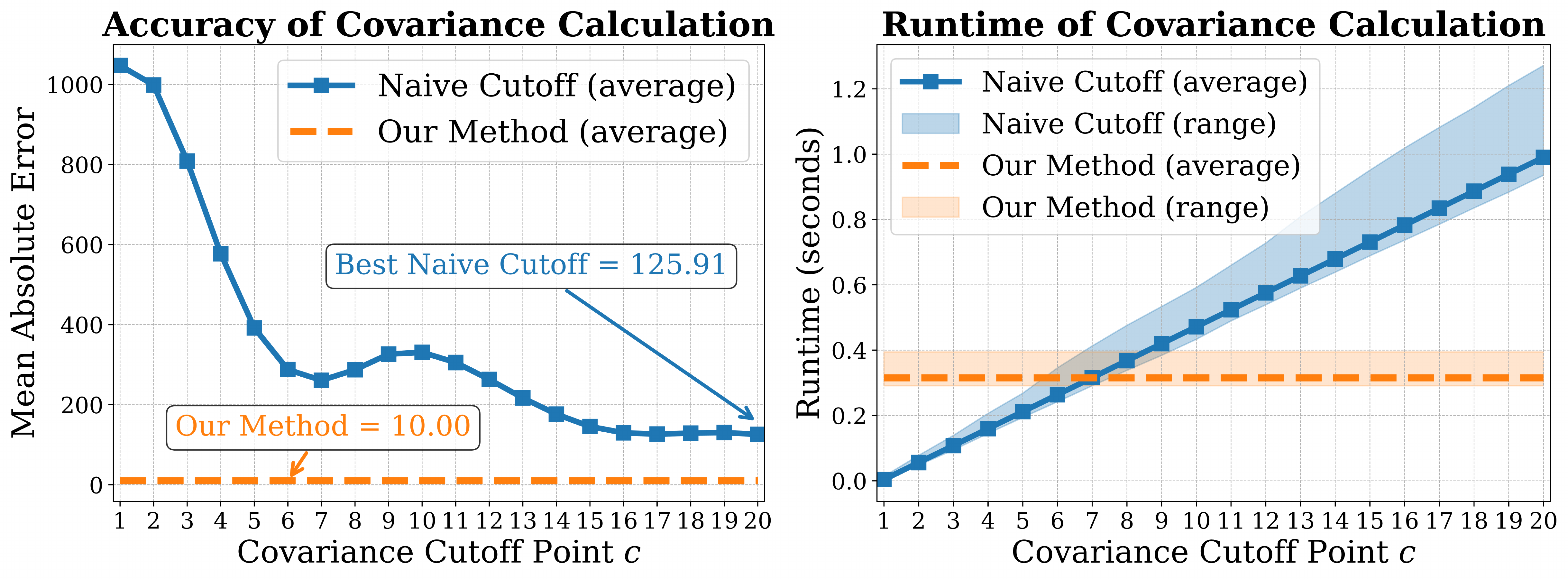}
    \caption{\textbf{Variance Prediction Accuracy and Efficiency.} 
    Our spectral truncation rule matches the ground-truth variance while significantly reducing computation compared with a large deterministic cutoff.}
    \label{fig:covariance_prediction}
\end{figure}

\vspace{0.5em}
\noindent\textbf{Validation.}
Across diverse $(S,B,t_d)$ settings, we compare our spectral truncation rule with deterministic summation of covariance terms. As shwon in \cref{fig:covariance_prediction}, in several challenging cases, even including 20 covariance terms fails to capture the long-range tail needed for accurate variance prediction, whereas our spectral rule recovers it precisely with negligible computational overhead. This demonstrates that a few dominant eigenmodes are sufficient to model the full correlation structure.

\subsection{Parallel Generation via Lookup Table and Delay Invariance}

Direct histogram synthesis becomes efficient once the photon-count 
statistics and temporal distribution are known, but two practical 
obstacles remain. First, accurate count prediction requires high temporal 
resolution (e.g., $2^{14}$ bins), where matrix operations and eigenvalue 
computations are expensive. Second, when synthesizing a full histogram 
cube, each pixel may have different $(S,B,\tau)$, and recomputing the 
statistics per pixel would undermine scalability.

A key observation resolves both issues. Although the reflectivity $S$ and 
background level $B$ affect the Markov--renewal behavior, the delay $\tau$ does not. Shifting the arrival flux by $\tau$ 
simply shifts the temporal PDF by the same amount, while the 
underlying renewal structure, and therefore the photon count statistics, remains unchanged. We formalize this property below.

\begin{theorem}[Delay Invariance]
\label{thm:delay_invariance}
Let $\mu_\tau$ and $\sigma_\tau^2$ denote the mean and variance of the
registered photon count, and let $\boldsymbol{\pi}_\tau$ denote the temporal PDF, under arrival rate
$\lambda_\tau(t) = \alpha\,s(t-\tau)+\lambda_b$.
Then for any delays $\tau,\tau'$,
\[
    \mu_\tau = \mu_{\tau'}, \quad
    \sigma_\tau^2 = \sigma_{\tau'}^2, \quad
    \boldsymbol{\pi}_{\tau'} = \text{shift}_{(\tau'-\tau)}\big(\boldsymbol{\pi}_\tau\big),
\]
i.e., the photon-count statistics are invariant to the delay and the
temporal distribution is only shifted in time.

\noindent (Proof provided in the Supplement.)
\end{theorem}

This invariance enables a simple and scalable mechanism for 
cube synthesis. We precompute the temporal 
PDF $\vpi$ and the effective mean $\mu$ and variance $\sigma^2$ of the Markov--renewal holding 
times- on a dense grid of $(S,B)$ while fixing the delay to a canonical 
value (e.g., the pulse centered in the period $t_r/2$). 

During synthesis, each pixel retrieves its $(S,B)$ entry from the lookup 
table, applies the appropriate temporal shift determined by $\tau$, and 
samples its histogram directly in a parallel fashion, as shown in \cref{alg:cube_sim}.
This eliminates per-pixel recomputation and enables accurate and 
efficient full-cube generation.

\begin{algorithm}[t]
\caption{Histogram Cube Generation}
\label{alg:cube_sim}
\small
\DontPrintSemicolon
\KwIn{System parameters $(t_r, N_r, \sigma_t, t_d, n_b)$, scene parameters $(\mZ, \mR, B)$, and realization $N_\mathrm{iter}$, where $\mZ,\mR \in \mathbb{R}^{H\times W}$ are depth and reflectivity images.}
\KwOut{$N_\mathrm{iter}$ histogram cubes $\mH_i \in \mathbb{R}^{H\times W\times n_b}$.}

\textbf{Step 1: Lookup.} Retrieve stationary distributions and MRP parameters:  
$(\mPi_0, \vmu_\mathrm{MRP}, \vsigma_\mathrm{MRP}^2) = \mathrm{Lookup}(\mR, B)$.\;

\textbf{Step 2: Temporal alignment.} Shift PDFs to account for scene-dependent delays:  
$\mPi = \mathrm{Shift}(\mPi_0, \mZ - t_r/2)$.\;

\textbf{Step 3: Photon count sampling.} Draw $N_\mathrm{iter}$ i.i.d. photon count maps:  
$\vn_i \sim \mathcal{N}(\vmu_\mathrm{MRP}, \vsigma_\mathrm{MRP}^2)$.\;

\textbf{Step 4: Histogram synthesis.} For each realization, sample histograms via  
$\mH_i \sim \mathrm{Multinomial}(\vn_i, \mPi)$.\;
\end{algorithm}

\section{Experiments}
\label{sec:experiments}

\begin{figure*}[t]
    \centering
    \includegraphics[width=\linewidth]{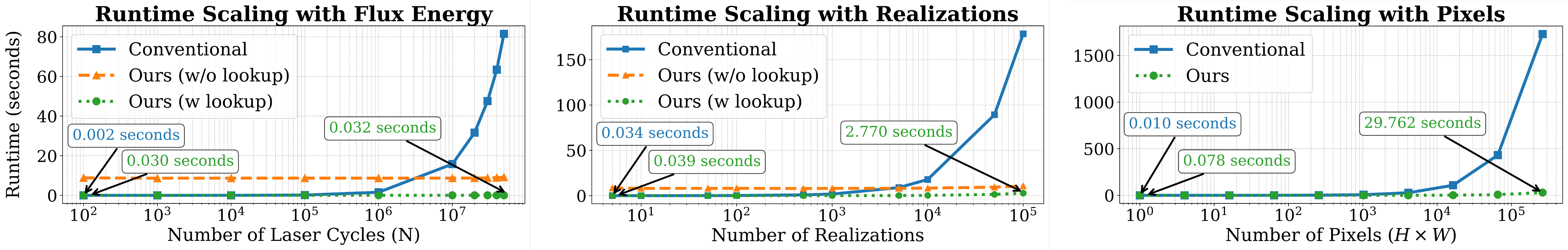}
    \caption{\textbf{Simulator Efficiency Comparison Analysis.} We compute the scaling of simulation runtime with flux, realizations, and pixels. MaRS with 
    the lookup table maintains near-constant runtime across all settings, 
    while MaRS without lookup and the conventional simulator slow down 
    significantly, demonstrating the benefit of the lookup-table design.}
    \label{fig:runtime_analysis}
\end{figure*}

We evaluate the proposed simulator along three key dimensions: accuracy, efficiency, and practical utility. We first verify its ability to reproduce photon count statistics under dead time, then measure runtime scaling with flux, realizations, and pixels, and finally assess its impact by training a reflectivity-estimation network using data generated by our model.

\subsection{Accuracy of Photon Count Prediction}
\label{subsec:sim_validation_accuracy}
We assess the accuracy of our photon count model by comparing its predicted
distribution to the empirical distribution obtained from the conventional
sequential simulator, treated as ground truth. Baselines include: 
(i) a renewal approximation that replaces the inhomogeneous flux with a 
constant-rate model of equal per-cycle energy, and 
(ii) Zhang et al.~\cite{zhang_2025_icip}.

\begin{figure}[t]
    \centering
    \includegraphics[width=\linewidth]{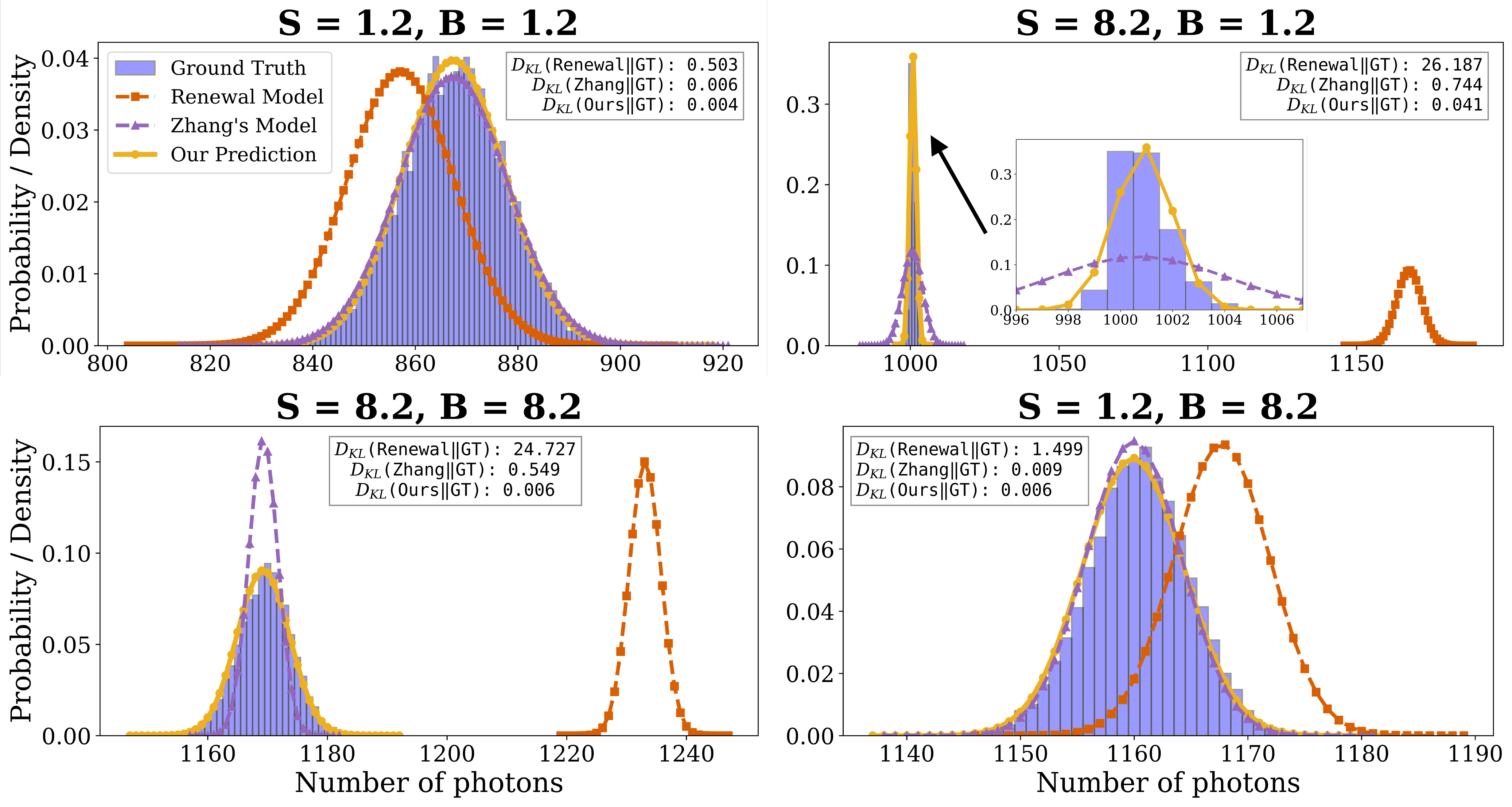}
    \caption{\textbf{Qualitative Count Prediction Accuracy.} We show the comparison of photon count distributions under several 
    representative $(S,B)$ combinations with $t_d = 75$\,ns. MaRS closely matches 
    the empirical ground truth in both shape and spread, whereas the renewal model 
    and Zhang~\cite{zhang_2025_icip} exhibit noticeable deviations.}
    \label{fig:count_qualitative}
\end{figure}

We evaluate a grid of $(S,B,t_r)$ values and summarize qualitative results
in Fig.~\ref{fig:count_qualitative} and quantitative metrics in 
Table~\ref{tab:metrics}. Figure~\ref{fig:count_qualitative} shows that the renewal model struggles to capture both the mean and variance, Zhang's model provides accurate mean 
estimates but mismatches variance under high signal levels, while MaRS achieves close matches robustly in various SBRs. Table~\ref{tab:metrics} reflects the same results under broader $(S, B, t_d)$ cases and evaluation metrics.

\begin{table}[t]
\centering
\resizebox{\columnwidth}{!}{%
\begin{tabular}{lcccc}
\toprule
\multirow{2}{*}{\textbf{Methods}} &
\multicolumn{2}{c}{\textbf{Distribution Metrics}} &
\multicolumn{2}{c}{\textbf{Moment Metrics}}\\
\cmidrule(lr){2-5}
 & \textbf{Wass Dist}~($\downarrow$) &
   \textbf{KL Div}~($\downarrow$) &
   \textbf{Mean Diff}~($\downarrow$) &
   \textbf{Var Diff}~($\downarrow$) \\
\midrule
Poisson  & 7966.130 & 24.208 & 7966.130 & 9257.464 \\
Renewal  & 207.565 & 18.821 & 207.564 &   45.927 \\
Zhang~\cite{zhang_2025_icip}  & 2.337 & 0.409 & 0.282 & 64.558 \\
\rowcolor{gray!25} \textbf{MaRS}  & \textbf{0.309} & \textbf{0.013} & \textbf{0.269} & \textbf{1.588} \\
\bottomrule
\end{tabular}%
}
    \caption{\textbf{Quantitative Accuracy of Photon Count Prediction.} Metrics are averaged over a grid of $(S,B,t_r)$ values. MaRS achieves 
    the closest match to the empirical ground truth in both distributional 
    (Wasserstein, KL) and moment (mean, variance) metrics, outperforming the 
    renewal approximation and Zhang~\cite{zhang_2025_icip}.}
\label{tab:metrics}
\vspace{-10pt}
\end{table}

\begin{table*}[t]
\centering
\newcommand{\highlightrow}{\cellcolor[HTML]{F8D7DA}}

\resizebox{\textwidth}{!}{%
\begin{tabular}{c|cccccccccccc}
\toprule[2pt]
\textbf{Simulator} &
\multicolumn{3}{c}{\textbf{SBR = $\mathbf{0.6}$}} &
\multicolumn{3}{c}{\textbf{SBR = $\mathbf{1.0}$}} &
\multicolumn{3}{c}{\textbf{SBR = $\mathbf{5.0}$}} &
\multicolumn{3}{c}{\textbf{SBR = $\mathbf{10.0}$}} \\
\cmidrule(lr){2-4}
\cmidrule(lr){5-7}
\cmidrule(lr){8-10}
\cmidrule(lr){11-13}
(Used for Training) & \textbf{PSNR $\uparrow$} & \textbf{SSIM $\uparrow$} & \textbf{LPIPS $\downarrow$} &
\textbf{PSNR $\uparrow$} & \textbf{SSIM $\uparrow$} & \textbf{LPIPS $\downarrow$} &
\textbf{PSNR $\uparrow$} & \textbf{SSIM $\uparrow$} & \textbf{LPIPS $\downarrow$} &
\textbf{PSNR $\uparrow$} & \textbf{SSIM $\uparrow$} & \textbf{LPIPS $\downarrow$} \\
\midrule[1.2pt]

\multirow{2}{*}{Zhang et al.~\cite{zhang_2025_icip}}
& $22.7444$ & $0.8908$ & $0.1220$ & $21.9506$ & $0.8820$ & $0.1128$ & $36.0575$ & $0.9599$ & $0.0506$ & $36.2273$ & $0.9652$ & $0.0481$ \\
& \highlightrow $22.5421$ & \highlightrow $0.8931$ & \highlightrow $0.1010$ &
  \highlightrow $22.2137$ & \highlightrow $0.9036$ & \highlightrow $0.0894$ &
  \highlightrow $36.1311$ & \highlightrow $0.9675$ & \highlightrow $0.0366$ &
  \highlightrow $36.3771$ & \highlightrow $0.9664$ & \highlightrow $0.0354$ \\
\midrule

\multirow{2}{*}{Renewal}
& $29.7230$ & $0.9192$ & $0.0544$ & $30.9704$ & $0.9329$ & $0.0413$ & $36.9140$ & $0.9611$ & $0.0347$ & $37.9956$ & $0.9802$ & $0.0272$ \\
& \highlightrow $12.1210$ & \highlightrow $0.6033$ & \highlightrow $0.3030$ &
  \highlightrow $11.6988$ & \highlightrow $0.6279$ & \highlightrow $0.2903$ &
  \highlightrow $14.7609$ & \highlightrow $0.7655$ & \highlightrow $0.2011$ &
  \highlightrow $15.2909$ & \highlightrow $0.8214$ & \highlightrow $0.1680$ \\
\midrule

\multirow{2}{*}{Poisson}
& $36.4946$ & $0.9668$ & $0.0225$ & $33.6827$ & $0.9616$ & $0.0231$ & $44.4941$ & $0.9888$ & $0.0131$ & $45.2165$ & $0.9901$ & $0.0126$ \\
& \highlightrow $7.0237$ & \highlightrow $-0.0753$ & \highlightrow $0.5160$ &
  \highlightrow $7.2490$ & \highlightrow $-0.0145$ & \highlightrow $0.5056$ &
  \highlightrow $8.5610$ & \highlightrow $0.3701$ & \highlightrow $0.4020$ &
  \highlightrow $8.7193$ & \highlightrow $0.4064$ & \highlightrow $0.3978$ \\
\midrule

\multirow{2}{*}{\textbf{MaRS}}
& $24.0321$ & $0.8893$ & $0.1315$ & $24.2821$ & $0.8946$ & $0.1063$ & $36.5793$ & $0.9705$ & $0.0458$ & $36.4003$ & $0.9722$ & $0.0431$ \\
& \highlightrow $\mathbf{24.2488}$ & \highlightrow $\mathbf{0.9084}$ & \highlightrow $\mathbf{0.1159}$ &
  \highlightrow $\mathbf{23.5776}$ & \highlightrow $\mathbf{0.9113}$ & \highlightrow $\mathbf{0.0899}$ &
  \highlightrow $\mathbf{36.9205}$ & \highlightrow $\mathbf{0.9748}$ & \highlightrow $\mathbf{0.0348}$ &
  \highlightrow $\mathbf{37.2704}$ & \highlightrow $\mathbf{0.9688}$ & \highlightrow $\mathbf{0.0366}$ \\
\bottomrule[2pt]
\end{tabular}%
}
\caption{
\textbf{Reflectivity Estimation Quantitive Results Across Simulators.} 
Quantitative results (PSNR, SSIM, LPIPS) across simulator domains and SBR levels. 
Unhighlighted rows show performance under the \emph{same-simulator} setup, 
while \textcolor[HTML]{E39A9A}{\textbf{highlighted rows}} indicate testing on the Gold-Standard (GS) simulator (\emph{cross-simulator} setup). 
The comparison shows how well each simulator generalizes to the real-world domain. 
Our approach achieves near-real-world fidelity with high efficiency.}
\label{tab:reconstruction_results}
\end{table*}

\begin{figure*}[t]
\centering
\includegraphics[width=\linewidth]{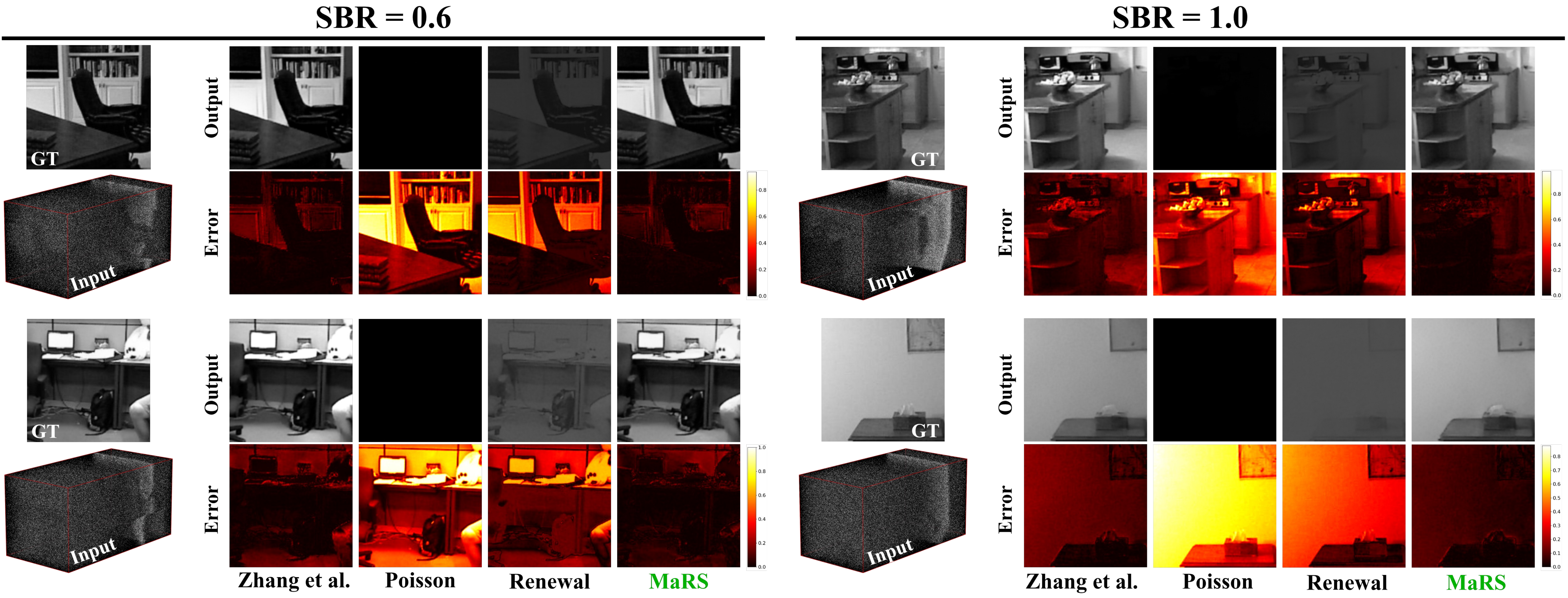}
\caption{\textbf{Reflectivity Estimation Comparison across Simulators.}
Reconstructed intensity accompanied by normalized error heatmaps for models trained under different simulators. 
Poisson shows strong artifacts from ignored dead time; Renewal improves but fails under varying flux; Zhang et al.~\cite{zhang_2025_icip} degrades at high SBR due to variance errors. 
\textbf{MaRS} best matches the GS reference, producing the most faithful reflectivity maps.}
\label{fig:experiments_visual_results}
\end{figure*}

\subsection{Efficiency of MaRS}
We benchmark the efficiency of MaRS by measuring runtime scaling with flux level, number of realizations, and number of pixels, as shown in \cref{fig:runtime_analysis}. All experiments are conducted in NumPy on a CPU machine equipped with an AMD Ryzen™ 7 9700X (8 cores / 16 threads) and 32GB DDR5 RAM.

For flux and realization scaling, we use the single-pixel simulator and compare three variants: the conventional sequential simulator, MaRS without the lookup table (requiring large matrix computations on the fly), and MaRS with the lookup table. For pixel scaling, we compare MaRS with lookup against the conventional simulator.

Across all experiments, MaRS with the lookup table exhibits near-constant runtime, as all statistics are retrieved from precomputed entries regardless of flux, realizations, or pixel count. MaRS without the lookup table degrades due to repeated high-resolution matrix operations, and the conventional sequential simulator scales poorly in all cases. These results confirm the effectiveness of the lookup-table design in enabling fast, scalable histogram cube synthesis.

\subsection{Downstream Task: Reflectivity Estimation}

A key goal of this work is to validate our per-pixel photon count model. Since reflectivity is proportional to the measured photon counts $N(t)$ for a fixed exposure time $t$, reflectivity estimation provides a direct test of how well the model captures true photon statistics.

\textbf{Implementation.}
We use a DDPM U-Net~\cite{ddpm}, modified to ingest the full spatio–temporal histogram cube (details in the supplementary). The network predicts the reflectivity map from this cube.

\textbf{Training and baselines.}
We train four versions of the same U-Net under different simulators: 
(1) Poisson, 
(2) Renewal + our temporal PDF estimator, 
(3) Zhang et al.~\cite{zhang_2025_icip} + our temporal PDF estimator, and 
(4) the proposed \textbf{MaRS}. 
For fairness, we replace competing temporal-PDF modules with ours wherever applicable.

\textbf{Evaluation.}
We test each model on data from its own simulator to confirm stable learning, then evaluate all models on the \textbf{Gold-Standard (GS)} simulator, which best reflects real sensor behavior. Better GS performance indicates more accurate photon-count and temporal-PDF modeling.

\textbf{Results (summary).}
Table~\ref{tab:reconstruction_results} and Fig.~\ref{fig:experiments_visual_results} show that \textbf{MaRS} provides the most accurate reflectivity estimates. Poisson fails due to ignored dead time. Renewal captures dead time but assumes constant flux, causing errors under time-varying illumination. Zhang et al.’s Gaussian model fits the mean but its variance error grows with SBR. MaRS models both mean and temporal variance correctly, producing the most faithful reconstructions.
\section{Conclusion}
\label{sec:conclusion}

SP-LiDAR simulation has long been split between two unsatisfying options: slow but accurate sequential simulators, or fast Poisson models that break under dead time and pile-up. Our work removes this trade-off.

We introduced an MRP-based simulator with three key components: 
(1) a direct Markov–Renewal formulation that models non-Markovian photon statistics and dead time analytically, 
(2) a spectral truncation scheme for fast and accurate covariance computation, and 
(3) a scalable synthesis pipeline that models the physics without relying on sequential sampling.

Our simulator matches the statistical behavior of the gold-standard sequential method while running over $6$ orders-of-magnitude faster—approaching Poisson-level speed without sacrificing physics. This removes the long-standing accuracy–speed compromise and enables fast, physically-correct histogram generation for next-generation learning-based SP-LiDAR algorithms.

\section{Acknowledgement}
The work is supported, in part, by the DARPA / SRC CogniSense JUMP 2.0 Center, NSF IIS-2133032, and NSF ECCS-2030570.

\newpage
{
    \small
    \bibliographystyle{ieeenat_fullname}
    \bibliography{main}
}

\clearpage

\maketitlesupplementary

\startcontents[supp] 
\printcontents[supp]{}{1}{\setcounter{tocdepth}{2}}



\section{Proofs of Theorems}
\label{suppsec:proof_thm}

We \emph{restate} each theorem in a formal manner and provide the corresponding proof. Equations reproduced from the main paper keep their original numbering, while new equations introduced in the supplement follow local numbering.

\subsection{Proof of Theorem~\textcolor{cvprblue}{1}}
\label{suppsubsec:proof_CLT}

\boxedthm{
\paragraph{Theorem 1 (CLT for $N(t)$ under Markov dependence).}
\textit{
    Consider a Markov–renewal process $\{(X_k, W_k)\}$ with arrival times $T_n = \sum_{k=1}^n W_k$ and counting process $N(t) = \max\{n : T_n \le t\}$. Assume the embedded chain $\{X_k\}$ is irreducible, recurrent, aperiodic, and geometrically mixing (spectral gap), and that $\E[W_k^2] < \infty$. Then,
    \begin{equation}
        \label{eq:CLT_Nt_supp}
        \frac{N(t) - \frac{t}{\mu}}{\sqrt{t\, \sigma^2/\mu^3}} \xrightarrow{d} \mathcal{N}(0,1), \qquad t\to\infty, \tag{4}
    \end{equation}
    where the effective mean and variance of the holding times $\{W_k\}$ are
    \begin{align}
        \mu
        & \bydef \lim_{n \rightarrow \infty} \frac{\E[T_n]}{n} = \lim_{n \rightarrow \infty} \frac{\E\left[\, \sum_{k=1}^n W_k \, \right]}{n}, \tag{5} \label{eq:equi_mean_def_supp} \\
        \sigma^2
        & \bydef \lim_{n \rightarrow \infty} \frac{\Var[T_n]}{n} = \lim_{n \rightarrow \infty} \frac{\Var\left[\, \sum_{k=1}^n W_k \, \right]}{n}. \tag{6} \label{eq:equi_var_def_supp}
    \end{align}
}
}

\noindent \textbf{Proof.}
Under the same assumptions of Theorem~\textcolor{cvprblue}{1}, the $n$th photon registration $T_n$, which is the \emph{sum} $T_n = \sum_{k=1}^n W_k$, satisfies
\begin{equation}
    \label{suppeq:CLT_Tn}
    \frac{T_n - n\mu}{\sqrt{n\sigma^2}} \xrightarrow{d} \mathcal{N}(0,1), \qquad n\to\infty.
\end{equation}

A rigorous proof can be found in the central limit theorems (CLTs) for additive functionals of ergodic Markov chains and regenerative/MRP settings \cite{Jones_2004_MarkovCLT,Komorowski_2012_CLT,Ferre_2012_L2Gap,Peligrad_2019_Additive}. In additional, we validate it empirically in Sec.~\textcolor{cvprblue}{6.1}. Intuition: $\{W_k\}_{k \in \mathbb{N}}$ are correlated but correlations decay geometrically by spectral gap; the variance inflation is considered in the effective variance $\sigma^2$.

To obtain \cref{eq:CLT_Nt_supp} from \cref{suppeq:CLT_Tn}, we need to invert the random variable $T_n$ to $N(t)$ via two equivalent events $\{N(t)\ge n\}\Leftrightarrow\{T_n\le t\}$. As a result,
\[
N(t) \xrightarrow{d} \mathcal{N}\left(\frac{t}{\mu},\frac{t \sigma^2}{\mu^3}\right), \qquad t\to\infty.
\]
This has been studied in the literature. See proof in classical results for renewal processes and Anscombe-type theorems \cite{Sigman_2009_Renewal,Gut_2011_Anscombe}. \hfill $\square$

\subsection{Proof of Theorem~\textcolor{cvprblue}{2}}
\label{suppsubsec:proof_effectiva_mean}

\boxedthm{
\paragraph{Theorem 2 (Effective mean of holding times $\mu$).}
\textit{
    Under the same assumptions of Theorem~\textcolor{cvprblue}{1}, the effective mean of holding times defined in \cref{eq:equi_mean_def_supp} becomes
    \begin{equation}
        \label{eq:equi_mean_supp}
        \mu = \vpi \mQ_{\vmu} \mathbf{1}, \tag{7}
    \end{equation}
    where $\vpi$ is the stationary temporal distribution of the embedded chain, $\mQ_{\vmu}\!=\mPtilde \odot \mM$ is the first moment of the transition kernel in Eq.~(\textcolor{cvprblue}{2}), where $\mPtilde$ is the state transition matrix and $\mM$ is the transition-dependent mean matrix of the holding times, as given in Lemma~\textcolor{cvprblue}{1}, and $\mathbf{1}$ is a column vector with all ones.
}
}

\noindent \textbf{Proof.} Starting from the definition in \cref{eq:equi_mean_def_supp},
\begin{align}
    \mu
    & = \lim_{n \rightarrow \infty} \frac{\E\left[\, \sum_{k=1}^n W_k \, \right]}{n} = \lim_{n \rightarrow \infty} \frac{\sum_{k=1}^n \E[W_k]}{n} \label{suppeq:equi_mean_expansion} \\
    & \overset{(i)}{\approx} \lim_{n \rightarrow \infty} \frac{\sum_{k=1}^n \E\left[\, \wss \, \right]}{n} = \E[\wss] \notag,
\end{align}
where $\wss$ represents the holding time after the system achieves the steady state and $(i)$ is based on the assumption that the system runs for long enough and converges to the steady state rapidly.

The expectation is over all possible transitions from any state $i$ to $j$. We apply the law of total expectation:
\begin{align}
    \mu
    & \overset{(i)}{=} \E_{i \rightarrow j}\bigl[\, \underset{\mu_{ij}}{\underbrace{\E[\, \wss \mid i \rightarrow j \, ]}} \, \bigr] \notag \\
    & \overset{(ii)}{=} \E_{X_k=i}\bigl[\, \E_{X_{k+1}=j}[\, \mu_{ij} \mid X_k=i \,]  \, \bigr] \notag \\
    & = \E_{X_k=i}\left[\, \textstyle\sum_j \Ptilde_{ij} \mu_{ij}  \, \right] \notag \\
    & = \textstyle\sum_i \pi_i \textstyle\sum_j \Ptilde_{ij} \mu_{ij} \notag \\
    & \overset{(iii)}{=} \vpi ( \underset{\mQ_{\vmu}}{\underbrace{\mPtilde \odot \mM}} ) \mathbf{1} = \E[\wss],
    \label{suppeq:equi_mean_mrp}
\end{align}
where $(i)$ sums over all possible transitions $i \to j$, $(ii)$ is based on summing over the possible next states $j$ first and then the possible current states $i$, and $(iii)$ is based on an equivalent representation of matrix forms. \hfill $\square$

\subsection{Proof of Theorem~\textcolor{cvprblue}{3} and Lemma~\textcolor{cvprblue}{2}}
\label{suppsubsec:proof_effective_var}

\boxedthm{
\paragraph{Theorem 3 (Effective variance of holding times $\sigma^2$).}
\textit{
    Under the same assumptions of Theorem~\textcolor{cvprblue}{1}, the effective variance of holding times defined in \cref{eq:equi_var_def_supp} is
    \begin{equation}
        \label{eq:equi_var_supp}
        \sigma^2 = \sigss^2 + 2 \gammass, \tag{8}
    \end{equation}
    which includes steady-state variance and covariance terms given by
    \begin{align}
        \sigss^2
        & = \vpi (\mQ_{\vmu^2} + \mQ_{\vsigma^2}) \mathbf{1}  - \mu^2, \label{eq:var_ss_supp} \tag{9} \\
        \gammass
        & = \sum_{l=1}^{\infty} \gamma_l \approx \sum_{l=1}^c \big(\vpi \mathbf{Q_{\vmu}} \mPtilde^{\, l-1} \mathbf{Q_{\vmu}} \mathbf{1} - \mu^2\big), \label{eq:covar_ss_supp} \tag{10}
    \end{align}
    respectively, where $c$ denotes a \textbf{cutoff} such that $\gamma_l \approx 0$ for $l > c$, $\vpi$ is the stationary distribution of the embedded chain, $\mQ_{\vmu^2}\!=\mPtilde \odot \mM \odot \mM$ and $\mQ_{\vsigma^2}\!=\mPtilde \odot \mSigma$ account for the second moment of the transition kernel in Eq.~(\textcolor{cvprblue}{2}), where $\mPtilde$ is the state transition matrix, $\mM$ and $\mSigma$ are transition-dependent mean and variance matrices of the holding times, as given in Lemma~\textcolor{cvprblue}{1}, $\mu$ is the effective mean provided in Theorem~\textcolor{cvprblue}{2}, and $\mathbf{1}$ is a column vector with all ones.
}
}

\noindent \textbf{Proof.} According to the definition in \cref{eq:equi_var_def_supp}, the equivalent variance of holding times $\sigma^2$ is the limiting variance of the latest absolute registration time $T_n$. If we expand $T_n$ into holding times as in \cref{suppeq:equi_mean_expansion}, one caveat is the necessity to consider nonzero covariances due to dependencies among state transitions. 

\begin{align}
    \Var[T_n]
    & = \Var\left[\, \textstyle\sum_{k=1}^n W_k \, \right] \notag \\
    & = \sum_{k=1}^n \Var[W_k] + 2 \sum_{\mathclap{1 \leq p < q \leq n}} \Cov[\, W_p, W_q \,] \notag \\
    & \overset{(i)}{\approx} \underset{= n \sigss^2}{\underbrace{\sum_{k=1}^n \underset{\coloneqq \sigss^2}{\underbrace{\Var[\wss]}}}}
    + 2 \biggl\{ (n-1) 
    \underset{\coloneqq \gamma_1}{\underbrace{\Cov[\, W_{\mathrm{ss}}, W_{\mathrm{ss}+1} \,]}} \notag \\
    & \qquad\qquad\qquad\qquad + (n-2) \Cov[\, W_{\mathrm{ss}}, W_{\mathrm{ss}+2} \,] \notag \\
    & \qquad\qquad\qquad\qquad + \Cov[\, W_{\mathrm{ss}}, W_{\mathrm{ss}+n-1} \,] \biggr\} \notag \\
    & = n \sigss^2 + 2 \sum_{l=1}^{n-1} (n-l) \gamma_l,
    \label{suppeq:equi_var_tn_derivation}
\end{align}
where $(i)$ is based on the assumption that the system runs for a long time and converges to its steady state quickly, and $\gamma_l = \Cov[\, W_{\mathrm{ss}}, W_{\mathrm{ss}+l} \,]$ represents the lag-$l$ covariance when the system is stationary.

To obtain the final equivalent variance $\sigma^2$, we deal with the steady-state variance $\sigss^2$ and the covariance series of $\gamma_l$, included in \cref{suppeq:equi_var_tn_derivation}, individually.

\noindent\textit{(1). Steady-state variance of holding times $\sigss^2$:}
\begin{align}
    \sigss^2
    & = \Var[W_{\mathrm{ss}}] = \E \left[W_{\mathrm{ss}}^2\right] - \left(\E[W_{\mathrm{ss}}]\right)^2 \notag \\
    & = \E_{i \rightarrow j}\left[\, \E[\, W_{\mathrm{ss}}^2 \mid i \rightarrow j \, ] \, \right] - \muss^2 \notag \\
    & = \E_{i \rightarrow j}\biggl[
        \bigl(
            \underset{\mu_{ij}}{\underbrace{\E[\, W_{\mathrm{ss}} \mid i \rightarrow j \, ]}} 
        \bigr)^2
        \!+\! \underset{\sigma_{ij}^2}{\underbrace{\Var[\, W_{\mathrm{ss}} \mid i \rightarrow j \, ]}}
    \biggr] \!-\! \mu^2 \notag \\
    & = \textstyle\sum_i \pi_i \textstyle\sum_j \Ptilde_{ij} \mu_{ij}^2 + \textstyle\sum_i \pi_i \textstyle\sum_j \Ptilde_{ij} \sigma_{ij}^2 - \mu^2 \notag \\
    & = \vpi ( \underset{\mQ_{\vmu^2}}{\underbrace{\mPtilde \odot \mM \odot \mM}} ) \mathbf{1} + \vpi ( \underset{\mQ_{\vsigma^2}}{\underbrace{\mPtilde \odot \mSigma}} ) \mathbf{1} - \mu^2. \label{suppeq:var_ss_expression}
\end{align}

\noindent\textit{(2). Steady-state lag-$l$ covariance $\gamma_l$:} We first look at the lag-$1$ covariance $\gamma_1$ and then extend it to the lag-$l$ one $\gamma_l$.
\begin{align}
    \gamma_1
    & = \Cov[\, W_{\mathrm{ss}}, W_{\mathrm{ss}+1} \,] \notag \\
    & = \E[\, W_{\mathrm{ss}}W_{\mathrm{ss}+1} \,] - \E[W_{\mathrm{ss}}] \E[W_{\mathrm{ss+1}}] \notag \\
    & \overset{(i)}{=} \E_{i \rightarrow j \rightarrow k}\left[\, \E[\, W_{\mathrm{ss}}W_{\mathrm{ss+1}} \mid i \rightarrow j \rightarrow k \, ] \, \right] - \mu^2 \notag \\
    & \overset{(ii)}{=} \E_{i \rightarrow j \rightarrow k}\Bigl[\, \E[\, W_{\mathrm{ss}} \mid i \rightarrow j \rightarrow k \, ] \notag \\
    & \qquad\qquad\qquad\quad\E[\, W_{\mathrm{ss+1}} \mid i \rightarrow j \rightarrow k \, ] \, \Bigr] - \mu^2 \notag \\
    & = \E_{i \rightarrow j \rightarrow k}\left[\, \mu_{ij} \mu_{jk} \, \right] - \mu^2 \notag \\
    & = \textstyle\sum_i \pi_i \textstyle\sum_j \Ptilde_{ij} \textstyle\sum_k \Ptilde_{jk} \mu_{ij} \mu_{jk} - \mu^2 \notag \\
    & = \textstyle\sum_i \pi_i \textstyle\sum_j \Ptilde_{ij} \mu_{ij} \textstyle\sum_k \Ptilde_{jk} \mu_{jk} - \mu^2 \notag \\
    & = \vpi \mathbf{Q_{\vmu}} \mathbf{Q_{\vmu}} \mathbf{1} - \mu^2 = \vpi \mathbf{Q_{\vmu}^\text{$2$}} \mathbf{1} - \mu^2, \notag
\end{align}
where $(i)$ changes from a one-step transition as in \cref{suppeq:equi_mean_mrp,suppeq:var_ss_expression} to two consectutive transitions and $(ii)$ is based on the conditional independence of $W_{\mathrm{ss}}$ and $W_{\mathrm{ss+1}}$, given the two-step transition $i \rightarrow j \rightarrow k$, as the waiting time only depends on its current state.

Similarly, we derive for the lag-$l$ covariance $\gamma_l$:
\begin{align}
    \gamma_l
    & = \Cov[\, W_{\mathrm{ss}}, W_{\mathrm{ss}+l} \,] \notag \\
    & = \E_{i \rightarrow j_1 \rightarrow \cdots \rightarrow j_l \rightarrow k}\left[\, \mu_{ij_1} \mu_{j_lk} \, \right] - \mu^2 \notag \\
    & = \textstyle\sum_i \pi_i \textstyle\sum_{j_1} \Ptilde_{ij_1} \mu_{ij_1} \textstyle\sum_{j_2} \Ptilde_{j_1j_2} \cdots \notag \\
    & \qquad\qquad\qquad\textstyle\sum_{j_l} \Ptilde_{j_{l-1}j_l} \textstyle\sum_k \Ptilde_{j_lk} \mu_{j_lk} - \mu^2 \notag \\
    & = \vpi \mathbf{Q_{\vmu}} \underset{(l-1) \mPtilde}{\underbrace{\mPtilde \cdots \mPtilde}} \mathbf{Q_{\vmu}} \mathbf{1} - \mu^2 \notag \\
    & = \vpi \mathbf{Q_{\vmu}} \mPtilde^{l-1} \mathbf{Q_{\vmu}} \mathbf{1} - \mu^2. \label{suppeq:cov_l_expression}
\end{align}

Although infinitely many covariance terms might be included in \cref{suppeq:equi_var_tn_derivation}, as $n \rightarrow \infty$, an interesting question arises: \textit{do all covariance terms matter?} We answer this in the following lemma.

\boxedeg{
\begin{lemma}[Convergence of $\gamma_l$]
    \label{lemma:covar_convergence}
    As $l$ grows, $\gamma_l$ will geometrically converge to $0$.

    \noindent (See proof in \cref{suppsubsec:proof_covar_convergence}).
\end{lemma}
}

To obtain the effective variance $\sigma^2$, we substitute \cref{suppeq:equi_var_tn_derivation,suppeq:cov_l_expression} into the definition of $\sigma^2$ in \cref{eq:equi_var_def_supp}:
\begin{align}
    \sigma^2
    & = \lim_{n \rightarrow \infty} \frac{\Var[T_n]}{n} \notag \\
    & = \lim_{n \rightarrow \infty} \frac{1}{n} \left[ \, n \sigss^2 + 2 \, \sum_{l=1}^{n-1} (n-l) \gamma_l \, \right] \notag \\
    & \overset{(i)}{\approx} \lim_{n \rightarrow \infty} \frac{1}{n} \left[ \, n \sigss^2 + 2 \, \sum_{l=1}^{c} (n-l) \gamma_l \, \right] \notag \\
    & = \sigss^2 +  2 \, \sum_{l=1}^{c} \cancelto{1}{\lim_{n \rightarrow \infty} \frac{n-l}{n}} \gamma_l \label{suppeq:equi_var} \\
    & \overset{(ii)}{=} \sigss^2 +  2 \, \underset{\coloneqq \gammass}{\underbrace{\sum_{l=1}^{\infty} \gamma_l}}, \notag
\end{align}
where $(i)$ is based on the assumption that $\gamma_c$ is the \emph{cutoff} covariance term, i.e. $\gamma_l \approx 0$ for $l > c$, and $n \gg c$, and $(ii)$ adds back higher-order covariance terms and defines the steady-state covariance $\gammass$. The proof is complete after plugging \cref{suppeq:cov_l_expression} into \cref{suppeq:equi_var}.
\hfill $\square$

\subsection{Proof of Theorem~\textcolor{cvprblue}{4}}
\label{suppsubsec:proof_delay_invariance}

\boxedthm{
\paragraph{Theorem 4 (Delay Invariance of Photon Registration Statistics).}
\textit{
Consider a photon arrival process with rate $\lambda_\tau(t) = \alpha\, s(t-\tau) + \lambda_b$, where $s(\cdot)$ is a Gaussian waveform with compact support, $\alpha > 0$ is the signal strength, $\lambda_b$ is a constant background rate, and $\tau$ is an arbitrary delay.
Assume a nonparalyzable detector with fixed dead time $t_d$ and asynchronous acquisition. Let $\mu_\tau$ and $\sigma_\tau^2$ denote the mean and variance of the registered photon count, and let $\boldsymbol{\pi}_\tau$ denote the temporal PDF of registered photons.
Then for any delays $\tau,\tau'$,
\[
    \mu_\tau = \mu_{\tau'}, \quad
    \sigma_\tau^2 = \sigma_{\tau'}^2, \quad
    \boldsymbol{\pi}_{\tau'} = 
    \mathrm{shift}_{\,(\tau'-\tau)}\!\left(\boldsymbol{\pi}_\tau\right).
\]
In other words, the count statistics are invariant to the delay, and the temporal PDF undergoes only a rigid shift.
}
}

\noindent \textbf{Proof.}
We divide the proof into two steps: (1) revisiting how the transition matrix $\mPtilde$, the core of the photon registration statistics, is constructed from the input parameters; (2) evaluating the effect of varying the delay $\tau$ on $\mPtilde$ and hence the statistics $\mu$, $\sigma^2$, and $\vpi$.

\noindent\textit{(1). Revisiting the construction of $\mPtilde$:} Following the notations and definitions in Sec.~\textcolor{cvprblue}{3.1}, the photon arrival flux function can be rewritten as
\[
\lambda(t) = S\,\calN(t-\tau) + \frac{B}{t_r}, \quad 0 \leq t < t_r.
\]
The shape of this waveform does not change with the delay $\tau$ if a finite support of the Gaussian laser pulse is assumed.

The cumulative flux function is defined as
\begin{align}
    F(t)
    & \coloneq \int_0^t \lambda(x)\, dx = S \int_0^t \calN(x; \tau, \sigma_t^2) \, dx + \frac{B}{t_r}t \notag \\
    & = S \left[\Phi\left(\frac{t - \tau}{\sigma_t}\right) - \Phi\left(-\frac{\tau}{\sigma_t}\right) \right] + \frac{B}{t_r}t,
    \label{suppeq:cumulative_flux}
\end{align}
where $\Phi(\cdot)$ is the cumulative density function (CDF) of the standard normal distribution.

The Markov model adopted to build the transition matrix $\mPtilde$ requires a discretization of the continuous time $t$ to finite states $\mathcal{S} = \{s_1,\dots,s_{n_b}\}$~\cite{Rapp_2019_Dead,zhang_2025_PDFacceleration}, where $n_b$ is the number of TCSPC bins. Denote the bin width by $\Delta = t_r / n_b$ and a vector of states by $\mathbf{s} = (s_1, s_2, \ldots, s_{n_b})^{\top}$. According to Zhang \etal~\cite{zhang_2025_PDFacceleration}, the transition matrix $\mPtilde$ results from a row normalization of the raw weight matrix $\mP$ that is given by
\begin{equation}
    \mP = (\mathbf{1}\,\vlambda^\top) \odot (\mJ_d \mE_{\mathrm{base}}),
    \label{suppeq:weight_matrix}
\end{equation}
where $\vlambda_j = \lambda(s_j)$, $\mJ_d$ is a permutation matrix that circularly rolls matrix rows by $d=\lceil x_d/\Delta \rceil$ according to the relative dead time $x_d = t_d \bmod t_r$, and the $(i, j)$th element of the base exponent matrix $\mE_{\mathrm{base}}$ is
\begin{equation}
    E_{ij} = \exp\left[- \Lambda \cdot \ind{s_i  > s_j} - F(s_j) + F(s_i) \right],
    \label{suppeq:exponent_matrix}
\end{equation}
where $\Lambda = S + B$ is the flux energy in one repetition period. Normalizing rows of $\mP$, we get the transition matrix $\mPtilde$. The stationary temporal PDF $\vpi$ is obtained by solving the left eigenvector corresponding to the largest eigenvalue, which is $1$ because $\mPtilde$ is a stochastic matrix, i.e. $\vpi \mPtilde = \vpi$, and $\vpi$ is \emph{unique}~\cite{Rapp_2019_Dead,zhang_2025_PDFacceleration}.

\noindent\textit{(2). Effect of the delay $\tau$:} Setting up enough prerequisites, we evaluate the impact of $\tau$. \textbf{For simplicity, we do not change the parameters associated with $\tau$ and denote those associated with $\tau'$ by a superscript $'$.} 

For any $\tau$ and $\tau'$, assume the difference of $\tau'$ and $\tau$ in the discrete space is $k = \lceil (\tau' - \tau) / \Delta \rceil$. Then, $\vlambda' = \mS_k \vlambda$, where $\mS_k$ is a permutation matrix that satisfies $\mS_k \mS_k^\top = \mI$, and left-multiplying $\mS_k$ corresponds to matrix row shifting by $k$ while right-multiplying $\mS_k^\top$ corresponds to matrix column shifting by $k$. Here, a positive $k$ value denotes downward row shifting and rightward column shifting, respectively, and vice versa.

For delay $\tau$, we plug \cref{suppeq:cumulative_flux} into \cref{suppeq:exponent_matrix}, and $E_{ij}$ becomes
\begin{align}
    E_{ij}
    & = \exp\left[- \Lambda \cdot \ind{s_i  > s_j} - F'(s_j) + F'(s_i) \right] \notag \\
    & = \underset{\coloneq A}{\underbrace{\exp\left[- \Lambda \cdot \ind{s_i  > s_j} \right] \, \exp\left[ \frac{B}{t_r}(s_i - s_j) \right]}} \notag \\
    & \qquad \ \exp\left\{ S \left[\Phi\left(\frac{s_i - \tau}{\sigma_t}\right) - \Phi\left(\frac{s_j - \tau}{\sigma_t}\right) \right] \right\} \notag \\
    & = A \exp\left\{ S \left[\Phi\left(\frac{s_i - \tau}{\sigma_t}\right) - \Phi\left(\frac{s_j - \tau}{\sigma_t}\right) \right] \right\},
    \label{suppeq:exponent_matrix_derivation}
\end{align}
where $A$ is a constant independent of $\tau$. For delay $\tau'$, similarly, its $(i,j)$th element of $\mE_{\mathrm{base}}'$ is
\begin{align}
    E_{ij}'
    & = A \exp\left\{ S \left[\Phi\left(\frac{s_i - \tau'}{\sigma_t}\right) - \Phi\left(\frac{s_j - \tau'}{\sigma_t}\right) \right] \right\} \notag \\
    & = A \exp\biggl\{ S \biggl[\Phi\biggl(\frac{(s_i + \tau - \tau') - \tau}{\sigma_t}\biggr) \notag \\
    & \qquad\qquad\qquad\ - \Phi\left(\frac{(s_j + \tau - \tau') - \tau}{\sigma_t}\right) \biggr] \biggr\}.
    \label{suppeq:exponent_matrix_prime}
\end{align}
Comparing \cref{suppeq:exponent_matrix_derivation,suppeq:exponent_matrix_prime}, the effect of the delay is equivalently an rearrangement of both the row and column indices of $\mE_{\mathrm{base}}$. In other words, we conclude that
\begin{equation}
    \mE_{\mathrm{base}}' = \mS_k \mE_{\mathrm{base}} \mS_k^\top.
    \label{suppeq:shift_of_exponent_matrix}
\end{equation}

The raw weight matrix $\mP'$ for the delay $\tau'$ is
\begin{align}
    \mP'
    & = (\mathbf{1}\,\vlambda'^\top) \odot (\mJ_d \mE_{\mathrm{base}}') \notag \\
    & \overset{(i)}{=} (\mathbf{1}\,\vlambda^\top \mS_k^\top) \odot (\mJ_d \mS_k \mE_{\mathrm{base}} \mS_k^\top) \notag \\
    &  \overset{(ii)}{=} (\mS_k \mathbf{1}\,\vlambda^\top \mS_k^\top) \odot (\mS_k \mJ_d \mE_{\mathrm{base}} \mS_k^\top) \notag \\
    &  \overset{(iii)}{=} \mS_k \underset{\mP}{\underbrace{\bigl[ (\mathbf{1}\,\vlambda^\top) \odot (\mJ_d \mE_{\mathrm{base}}) \bigr]}} \mS_k^\top,
    \label{suppeq:shift_of_weight_matrix}
\end{align}
where $(i)$ is based on the substitutions of $\vlambda'$ and \cref{suppeq:shift_of_exponent_matrix}, $(ii)$ is based on the facts that any row shift of $\mathbf{1}$ stays the same vector and row shifting operations are interchangeable, and $(iii)$ is based on the fact that the same permutation operations and the element-wise multiplication can be exchanged.

Since the row normalization from the raw weight matrix $\mP$ to the transition matrix $\mPtilde$ preserves the row and column shifting relationship and $\mS_k \mS_k^\top = \mI$ holds, we have
\begin{equation}
    \mP = \mS_k^\top \mP' \mS_k, \quad \mPtilde = \mS_k^\top \mPtilde' \mS_k.
    \label{suppeq:shift_of_transition_matrix}
\end{equation}
Plugging \cref{suppeq:shift_of_transition_matrix} into $\vpi \mPtilde = \vpi$, we obtain
\begin{equation}
    \vpi \mS_k^\top \mPtilde' \mS_k  = \vpi, \ \Rightarrow \
    \vpi \mS_k^\top \mPtilde'  = \vpi \mS_k^\top. \notag
\end{equation}
Since $\vpi'$ is the unique solution to $\vpi' \mPtilde' = \vpi'$, it holds that
\begin{equation}
    \vpi'  = \vpi \mS_k^\top.
    \label{suppeq:shift_of_pi}
\end{equation}
Alternatively, the delay only causes a corresponding column shift in the row vector $\vpi$, i.e. $\boldsymbol{\pi}_{\tau'} = \mathrm{shift}_{\,(\tau'-\tau)}\!\left(\boldsymbol{\pi}_\tau\right)$.

Next, we extend the analysis to the count statistics
\[
    N(t) \sim \calN(\mu, \sigma^2), \quad \mu = \frac{t}{\mu_0}, \quad \sigma^2 = \frac{t \sigma_0^2}{\mu^3},
\]
where $\mu_0$ and $\sigma_0^2$ are the effective mean and variance of the holding times, respectively. From \cref{suppeq:equi_mean_mrp}, for delay $\tau$, \[
    \mu_0 = \vpi \mQ_{\vmu} \mathbf{1} = \vpi ( \mPtilde \odot \mM ) \mathbf{1}.
\]
Following a similar derivation as from \cref{suppeq:exponent_matrix_derivation,suppeq:exponent_matrix_prime,suppeq:shift_of_exponent_matrix}, for delay $\tau'$,
\[
    \mM' = \mS_k \mM \mS_k^\top, \quad \mQ_{\vmu}' = \mS_k \mQ_{\vmu} \mS_k^\top,
\]
and therefore
\begin{align*}
    \mu_0'
    & = \vpi' \mQ_{\vmu}' \mathbf{1} = (\vpi \mS_k^\top) (\mS_k \mQ_{\vmu} \mS_k^\top) \mathbf{1} \notag \\
    & = \vpi \mQ_{\vmu} \mS_k^\top \mathbf{1} \notag \\
    & \overset{(i)}{=} \vpi \mQ_{\vmu} \mathbf{1} = \mu_0,
\end{align*}
where $(i)$ is based on the fact that row shifting does not change the column vector $\mathbf{1}$. Then, for the same exposure time $t$ and any delay $\tau$ and $\tau'$, the means of the photon count stay the same, i.e.
\begin{equation}
    \mu' = \mu, \ \Leftrightarrow \ \mu_\tau = \mu_{\tau'}.
    \label{suppeq:shift_invariance_mean}
\end{equation}

Following similar analyses for $\mSigma$, $\sigss$, $\gammass$, and ultimately $\sigma_0^2$, for any delay $\tau$ and $\tau'$, the variances of the photon count are also the same, i.e.
\begin{equation}
    (\sigma^2)' = \sigma^2, \ \Leftrightarrow \ \sigma_\tau^2 = \sigma_{\tau'}^2.
    \label{suppeq:shift_invariance_var}
\end{equation}

Combining \cref{suppeq:shift_of_pi,suppeq:shift_invariance_mean,suppeq:shift_invariance_var}, the proof is complete.
\hfill $\square$

\section{Proofs of Additional Results}
\label{suppsec:proof_additional_results}

\subsection{Proof of Lemma~\textcolor{cvprblue}{1}}
\label{suppsubsec:proof_cond_mean_var_holding_time}

We rewrite Lemma~\textcolor{cvprblue}{1} in a more formal way as below.

\boxedeg{
\paragraph{Lemma 1 (Conditional Mean and Variance of the Holding Time).}
\textit{
Consider a nonparalyzable detector with dead time $t_d$ and repetition period $t_r$. Let $W$ denote the holding time between two consecutive registrations, and suppose the embedded chain makes a transition $i \rightarrow j$. Let $x_{i'} = (x_i + t_d) \bmod t_r$ be the relative
reactivation time, and define
\[
    t_{i'j} \bydef (x_j - x_{i'}) \bmod t_r
\]
as the minimum awake interval before the next registration.
Let $\Lambda$ be the incident flux energy in one period.
Then the conditional mean and variance of $W$ are
\[
    \mu_{ij} = t_d + t_{i'j} + \frac{e^{-\Lambda}}{1 - e^{-\Lambda}}\, t_r,
    \quad
    \sigma_{ij}^2 = \frac{e^{-\Lambda}}{(1 - e^{-\Lambda})^2}\, t_r^2.
\]
}
}

\begin{figure*}[t]
    \centering
    \includegraphics[width=\linewidth]{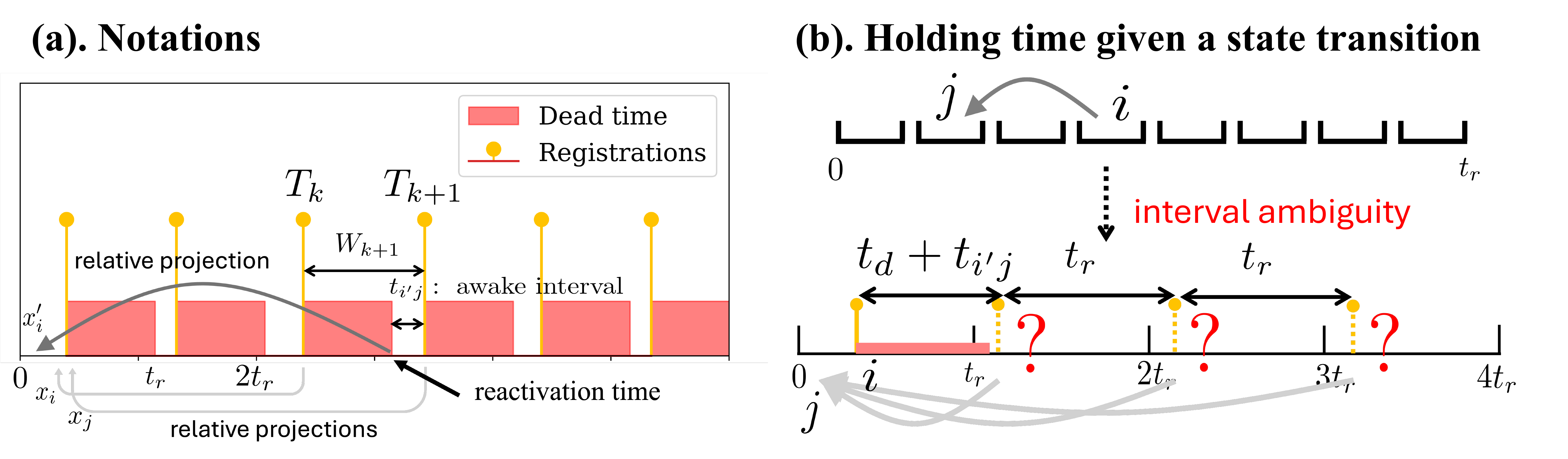}
    \caption{\textbf{Proof of Lemma 1.}
    \textbf{(a)} Summary of the timing geometry for a transition $i \!\rightarrow\! j$:  the detector reactivates at $x_{i'}$, the next registration occurs at $x_j$,  and the minimal awake interval is $t_{i'j}$.  \textbf{(b)} The absolute timestamps $T_k$ and $T_{k+1}$ differ by a deterministic offset $t_d + t_{i'j}$ plus a stochastic number of full repetition cycles, giving rise to the distribution of the holding time $W \mid i \!\rightarrow\! j$.
    }
    \label{suppfig:lemma1}
\end{figure*}

\noindent \textbf{Proof.}
We summarize the notations in \cref{suppfig:lemma1}(a). The high-level idea is that conditioned on a transition $i \rightarrow j$, i.e. the current state is $X_k = i$ and the next state is $X_{k+1} = j$, the \emph{relative} locations of two consecutive registrations are deterministic. However, an ambiguity still exists in the \emph{absolute} timestamps $T_k$ and $T_{k+1}$, between which the number of complete repetition cycles is stochastic. A longer waiting time corresponds to a larger number of full cycles, which occurs with exponentially decaying probability. This idea is illustrated in \cref{suppfig:lemma1}(b). Our goal is to derive a closed-form expression for $W \mid i \rightarrow j$, and then compute its conditional mean and variance.

We first describe the deterministic geometry of the holding time. If the current registration occurs at $T_k$ with an embedded state $x_i$, the detector becomes active again at $T_k + t_d$. The corresponding state of this reactivation point is $x_{i'} = (x_i + t_d) \bmod t_r$~\cite{zhang_2025_PDFacceleration}. Given a transition $i \rightarrow j$, the next registration occurs at $T_{k+1}$ with state $x_j$, and the \emph{minimum} awake interval between the reactivation and the next registration is  
\[
t_{i^{\prime}j}
    = \bigl(T_{k+1} - (T_k + t_d)\bigr) \bmod t_r
    = (x_j - x_{i^\prime}) \bmod t_r.
\]
This term is fully deterministic once the transition $i\!\rightarrow\!j$ is known. Considering the dead time, the deterministic offset in the holding time is $t_d + t_{i^{\prime}j}$.

In addition to this offset, the detector may wait through several full repetition periods before a photon is registered.  
Let  
\[
R \bydef \left\lfloor \frac{T_{k+1} - (T_k + t_d)}{t_r} \right\rfloor
\]
denote the number of complete periods between reactivation and the next registration.  
The waiting time is longer when the per-period arrival energy $\Lambda$ is small, and $R$ captures precisely this stochastic variability.

Combining the deterministic offset with the random number of whole periods, the support of $W \mid i \rightarrow j$ is  
\[
\left\{
    t_n :\;
    t_n = t_d + t_{i^{\prime}j} + n\, t_r,
    \; n \in \mathbb{N}^{+} \cup \{0\}
\right\}.
\]
Since the event that the next registration falls after $n$ full periods depends only on the arrival flux and not on the specific transition, the conditional PMF is
\begin{align}
    f_{ij}(t_n)
    &= \Pr(W = t_n \mid i \rightarrow j) \notag \\
    &= \Pr(R = n) \notag \\
    &= (1 - e^{-\Lambda})\, e^{-n\Lambda},
    \quad n \in \mathbb{N}^{+} \cup \{0\}, \notag
\end{align}
where the distribution of $R$ follows standard results for periodic arrival processes \cite{Rapp_2019_Dead}.

Next, we derive the mean of the conditional holding time $\mu_{ij}$:
\begin{align}
    \mu_{ij}
    & = \E[\, W \mid i \rightarrow j \, ] = \sum_{n=0}^{\infty} t_n f_{ij}(t_n) \notag \\
    & = \sum_{n=0}^{\infty} (t_d + t_{i^{\prime}j} + n t_r) (1 - e^{-\Lambda})e^{-n \Lambda} \notag \\
    & = (t_d + t_{i^{\prime}j}) \underset{ = 1}{\underbrace{\sum_{n=0}^{\infty} (1 - e^{-\Lambda})e^{-n \Lambda}}} \notag \\
    & \qquad\qquad\qquad\qquad + t_r (1 - e^{-\Lambda}) \sum_{n=0}^{\infty} n e^{-n \Lambda} \notag \\
    & = t_d + t_{i^{\prime}j} + t_r (1 - e^{-\Lambda}) e^{-\Lambda} \sum_{n=1}^{\infty} n e^{-(n-1) \Lambda} \notag \\
    & = t_d + t_{i^{\prime}j} + t_r (1 - e^{-\Lambda}) e^{-\Lambda} \frac{1}{(1 - e^{-\Lambda})^2} \notag \\
    & = t_d + t_{i^{\prime}j} + \underset{\coloneq K}{\underbrace{\frac{e^{-\Lambda}}{1 - e^{-\Lambda}}}} t_r,
    \label{suppeq:mu_ij}
\end{align}
where the deterministic $t_d + t_{i^{\prime}j}$ depends on the dead time duration and the state transition, while the term $K t_r$ characterizes the averaged waiting time across multiple repetition periods. The latter term approaches zero if the flux energy $\Lambda$ is large, since the system's active time between two registrations is unlikely to exceed one repetition period.

With the closed-form equation for $\mu_{ij}$, we derive the variance of the conditional holding time:
\begin{align}
    \sigma_{ij}^2
    & = \Var[\, W \mid i \rightarrow j \, ] \notag \\
    & = \E[\, (W - \mu_{ij})^2 \mid i \rightarrow j \, ] \notag \\
    & = \sum_{n=0}^{\infty} (t_n - \mu_{ij})^2 f_{ij}(t_n) \notag \\
    & = \sum_{n=0}^{\infty} (n - K)^2 t_r^2 f_{ij}(t_n) \notag \\
    & = \underset{\coloneqq L_0}{\underbrace{\sum_{n=0}^{\infty} n^2 t_r^2 f_{ij}(t_n)}} - \underset{\coloneqq L_1}{\underbrace{\sum_{n=0}^{\infty} 2nK t_r^2 f_{ij}(t_n)}} \notag \\
    & \qquad\qquad\qquad\qquad\qquad + \underset{\coloneqq L_2}{\underbrace{\sum_{n=0}^{\infty} K^2 t_r^2 f_{ij}(t_n)}},
    \label{suppeq:sigma2_ij_derivation}
\end{align}
where $L_0$ is given by
\begin{align}
    L_0
    & = \sum_{n=0}^{\infty} n^2 t_r^2 f_{ij}(t_n) \notag \\
    & = (1 - e^{-\Lambda}) e^{-\Lambda} t_r^2 \sum_{n=1}^{\infty} n^2 e^{-(n-1) \Lambda} \notag \\
    & = (1 - e^{-\Lambda}) e^{-\Lambda} t_r^2 \frac{1+e^{-\Lambda}}{(1 - e^{-\Lambda})^3} \notag \\
    & = \frac{e^{-\Lambda} (1+e^{-\Lambda})}{(1 - e^{-\Lambda})^2} t_r^2, \notag
\end{align}
$L_1$ is given by
\begin{align}
    L_1
    & = \sum_{n=0}^{\infty} 2nK t_r^2 f_{ij}(t_n) \notag \\
    & = 2K(1 - e^{-\Lambda}) e^{-\Lambda} t_r^2 \sum_{n=1}^{\infty} n e^{-(n-1) \Lambda} \notag \\
    & = 2K(1 - e^{-\Lambda}) e^{-\Lambda} t_r^2 \frac{1}{(1 - e^{-\Lambda})^2} \notag \\
    & = \frac{2K e^{-\Lambda}}{1 - e^{-\Lambda}} t_r^2 = 2K^2 t_r^2, \notag
\end{align}
and $L_2$ is given by
\begin{align}
    L_2
    & = \sum_{n=0}^{\infty} K^2 t_r^2 f_{ij}(t_n) \notag \\
    & = K^2 t_r^2 \sum_{n=0}^{\infty} f_{ij}(t_n) \notag \\
    & = K^2 t_r^2. \notag
\end{align}
Plugging $L_0$, $L_1$, and $L_2$ into \cref{suppeq:sigma2_ij_derivation},
\begin{align}
    \sigma_{ij}^2
    & = L_0 - L_1 + L_2 \notag \\
    & = \frac{e^{-\Lambda} (1+e^{-\Lambda})}{(1 - e^{-\Lambda})^2} t_r^2 - 2K^2 t_r^2 + K^2 t_r^2 \notag \\
    & = \left[ \frac{e^{-\Lambda} (1+e^{-\Lambda})}{(1 - e^{-\Lambda})^2} - \frac{e^{-2\Lambda}}{(1 - e^{-\Lambda})^2} \right] t_r^2 \notag \\
    & = \frac{e^{-\Lambda}}{(1 - e^{-\Lambda})^2} t_r^2,
    \label{suppeq:sigma2_ij}
\end{align}
which indicates that the conditional variance is a constant independent of the transition. Similarly, when $\Lambda$ increases, since incoming photons become denser, the variance is expected to approach zero.

Combining \cref{suppeq:mu_ij,suppeq:sigma2_ij}, the proof of Lemma~\textcolor{cvprblue}{1} is complete.
\hfill $\square$

\subsection{Proof of Lemma~\ref{lemma:covar_convergence}}
\label{suppsubsec:proof_covar_convergence}

Under the same assumptions of Theorem~\textcolor{cvprblue}{1}, the $n$-step transition matrix $\mPtilde^{l-1}$ will converge to a matrix with replicated and unique stationary distribution $\vpi$ as rows, i.e.
\[
\lim_{n \rightarrow \infty} \mPtilde^n = \mathbf{1} \vpi. 
\]
Then, the limiting covariance term is
\begin{align}
    \lim_{l \rightarrow \infty} \gamma_l
    & = \lim_{l \rightarrow \infty} \vpi \mathbf{Q_{\vmu}} \mPtilde^{l-1} \mathbf{Q_{\vmu}} \mathbf{1} - \mu^2 \notag \\
    & = \vpi \mathbf{Q_{\vmu}} \left(
        \lim_{l \rightarrow \infty} \mPtilde^{l-1}
    \right) \mathbf{Q_{\vmu}} \mathbf{1} - \mu^2 \notag \\
    & = \vpi \mathbf{Q_{\vmu}} \left( \mathbf{1} \vpi \right) \mathbf{Q_{\vmu}} \mathbf{1} - \mu^2 \notag \\
    & = \bigl(\underset{= \mu}{\underbrace{\vpi \mathbf{Q_{\vmu}} \mathbf{1}}} \bigr) \bigl(\underset{= \mu}{\underbrace{\vpi \mathbf{Q_{\vmu}} \mathbf{1}}} \bigr) - \mu^2 = 0. \notag
\end{align}
The convergence speed is geometric, governed by the magnitude of the few leading eigenvalues of $\mPtilde$~\cite{Rapp_2019_Dead,zhang_2025_PDFacceleration}.
\hfill $\square$

\section{MRP as a Unified Framework}

A common strategy for analyzing photon counting systems is to model:
(i) the inter-arrival times $\{W_k\}$,
(ii) the arrival times $T_n = \sum_{k=1}^n W_k$, and
(iii) the counting process $N(t) = \max\{n : T_n \le t\}$.  
This viewpoint reveals a natural hierarchy: once the law of $W_k$ is known, the distributions of $T_n$ and $N(t)$ follow.

Our Markov-renewal-process (MRP) formulation generalizes this hierarchy.  
It allows $W_k$ to depend on an embedded state sequence $\{X_k\}$, capturing correlations induced by flux variations or nontrivial timing geometry.  
Classical models arise as special cases:
the \emph{homogeneous Poisson process} corresponds to i.i.d.\ exponential $W_k$ (no state, no dead time);
the \emph{delayed renewal process} corresponds to i.i.d.\ $W_k$ with a deterministic offset from dead time (one-state MRP with no covariance).

The following subsections revisit these classical models from the MRP perspective, illustrating how they fit into the same unified framework.

\subsection{Special Case: Poisson}
Assuming a \emph{homogeneous Poisson process} and \emph{no dead time}, the photon registration process is equivalent to its preceding arrival process. The arrival flux is a constant $\lambda(t) = \lambda_0$, so all the inter-arrival times $W_k$ are i.i.d.\ exponential random variables with rate $\lambda_0$. The $n$th arrival time $T_n$ is an Erlang random variable with shape $n$ and rate $\lambda_0$. Consequently, the count $N(t)$ follows a Poisson distribution with mean $\lambda_0 t$.

\paragraph{From Exponential to Erlang.}
The PDF of the i.i.d.\ exponential random variable $W_i$ is
\[
    f_{W_i}(t) = \lambda_0 e^{-\lambda_0 t} , \quad t \geq 0.
\]

The PDF of $T_n = \sum_{i=1}^n W_i$ can be obtained by convolving $f_{W_i}(t)$ with itself for $n$ times. Alternatively, we can compute it in the Laplace domain:
\begin{align*}
    \mathcal{L}\{f_{W_i}(t)\}(s)
    & = \int_0^{\infty} e^{-st} (\lambda_0 e^{-\lambda_0 t}) \, dt \\
    & = \frac{\lambda_0}{\lambda_0 + s}, \quad s > -\lambda_0.
\end{align*}
Using the convolution theorem,
\[
\mathcal{L}\{f_{T_n}(t)\}(s) = \prod_{i=1}^n \frac{\lambda_0}{\lambda_0 + s} =  \left( \frac{\lambda_0}{\lambda_0 + s} \right)^n.
\]
From the known Laplace pair (when $s+a>0$)
\[
\mathcal{L}^{-1} \left\{ \frac{1}{(s + a)^n} \right\} = \frac{t^{n - 1} e^{-a t}}{(n - 1)!}, \quad t \geq 0,
\]
we compute the inverse Laplace transform and obtain
\begin{align*}
    f_{T_n}(t)
    & = \mathcal{L}^{-1} \left\{ \left( \frac{\lambda_0}{\lambda_0 + s} \right)^n \right\} = \lambda_0^n \cdot \mathcal{L}^{-1} \left\{ \frac{1}{(\lambda_0+s)^n} \right\} \\
    & = \lambda_0^n \cdot \frac{t^{n - 1} e^{-\lambda_0 t}}{(n - 1)!}, \quad t \geq 0
\end{align*}

This is the PDF of the Erlang distribution with shape parameter \( n \) and rate \( \lambda_0 \).

\paragraph{From Erlang to Poisson}
Using the two equivalent events $\{N(t) \geq n\} = \{T_n \leq t\}$, we copmpute
\begin{align}
    p_{N(t)}(n)
    & = \Pr(N(t) = n) \notag \\
    & = \Pr(N(t) \geq n) - \Pr(N(t) \geq n + 1) \notag \\
    & = \Pr(T_n \leq t) - \Pr(T_{n+1} \leq t) \notag \\
    & = F_{T_n}(t) - F_{T_{n+1}}(t),
    \label{suppeq:Poisson_rv}
\end{align}
where $F_{T_n}(t)$ is the cumulative density function (CDF) of $T_n$, which can be calculated from its PDF:
\[
F_{T_n}(t) = \int_0^t f(u)\,du = \int_0^t \frac{\lambda_0^n u^{n-1} e^{-\lambda_0 u}}{(n-1)!}\,du.
\]
This integral is equivalent to the regularized lower incomplete gamma function:
\[
F_{T_n}(t) = \frac{\gamma(n, \lambda_0 t)}{(n-1)!},
\]
where
\[
\gamma(n, x) = \int_0^x u^{n-1} e^{-u} du.
\]
When $n$ is a positive integer, this has a closed-form expression:
\[
\gamma(n, x) = (n - 1)! \left(1 - e^{-x} \sum_{k=0}^{n-1} \frac{x^k}{k!} \right).
\]
Thus, the Erlang CDF becomes:
\begin{equation}
    F_{T_n}(t) = 1 - \sum_{k=0}^{n-1} \frac{(\lambda_0 t)^k}{k!} e^{-\lambda_0 t}.
    \label{suppeq:Erlang_CDF}
\end{equation}
Plugging \cref{suppeq:Erlang_CDF} into \cref{suppeq:Poisson_rv},
\[
p_{N(t)}(n) = \frac{(\lambda_0 t)^n}{n!} e^{-\lambda_0 t}, \quad n \in \mathbb{N}^{+} \cup \{0\},
\]
i.e., $N(t) \sim \texttt{Poisson}(\lambda_0 t)$.

\subsection{Special Case: Renewal}
We now still assume a \emph{homogeneous Poisson process}, but a \emph{constant dead time}. Introducing a fixed dead time $t_d$ breaks the memoryless structure of the exponential distribution, but the inter-registration times remain i.i.d.\ after the first event.  
Each $W_k$ (for $k \ge 2$) consists of a deterministic inactive interval $t_d$ followed by an exponential wait with rate $\lambda_0$.  
Thus,
\[
\mu_\mathcal{R} = \E[W_k] = t_d + \frac{1}{\lambda_0}, 
\quad
\sigma_\mathcal{R}^2 = \Var[W_k] = \frac{1}{\lambda_0^2}.
\]

This leads to a delayed renewal process.  
Classical renewal theory implies that, asymptotically, $N(t)$ converges to a Gaussian~\cite{feller_1968_probability,Yu_2000_counting}, with mean and variance
\[
\E[N(t)] = \frac{t}{\mu_\mathcal{R}}, \quad
\Var[N(t)] = \frac{t \sigma_\mathcal{R}^2}{\mu_\mathcal{R}^3},
\]
so
\[
\E[N(t)] = \frac{\lambda_0 t}{1 + \lambda_0 t_d}, \quad
\Var[N(t)] = \frac{\lambda_0 t}{(1 + \lambda_0 t_d)^3}.
\]
Setting $t_d = 0$ recovers the Poisson moments, showing that the delayed renewal model is a strict generalization of the homogeneous Poisson model.

Moreover, this renewal model is precisely the one-state special case of our MRP
(only one state, no covariance terms).  
It corresponds to scenarios with constant flux or vanishing SBR, where transition structure becomes irrelevant. This explains why the prediction from the renewal model becomes better when the SBR reduces in Fig.~\textcolor{cvprblue}{5}.

\section{Spectral Analysis of Long-Lag Covariances}
\label{suppsec:spectral_analysis}

In this section, we detail our spectral truncation rule—its motivation, derivation, and practical usage

\subsection{Challenges of Long-Lag Covariances}
\begin{figure}[t]
    \centering
    \includegraphics[width=\linewidth]{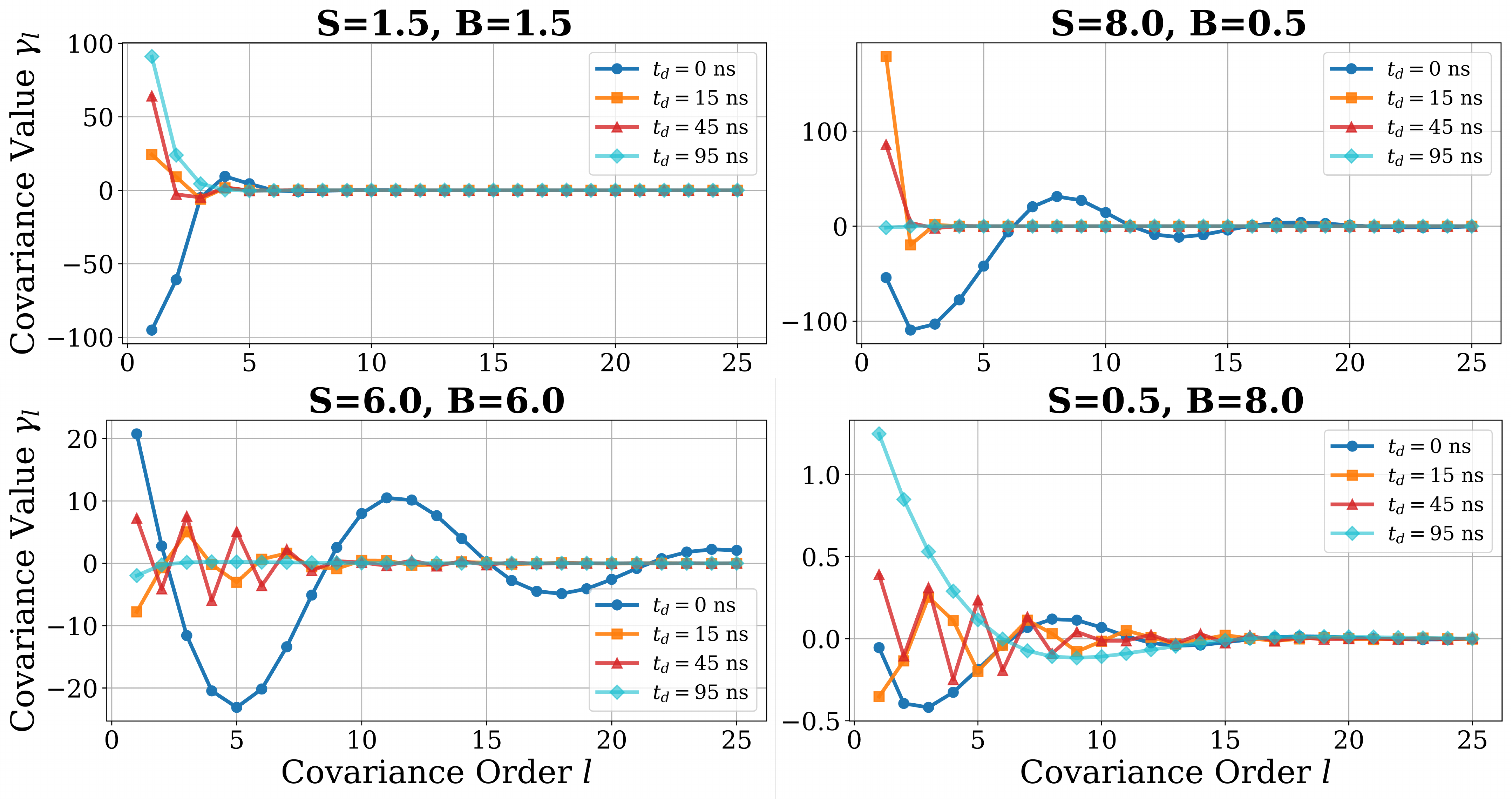}
    \caption{\textbf{Convergence Dynamics of $\{\gamma_l\}$ for Different $(S,B,t_d)$ Settings.} The diverse convergence rates and oscillations indicate the challenge of selecting a single cutoff $c$ for all cases.}
    \label{suppfig:convergence_dynamics}
\end{figure}

From the main paper, the variance of $N(t)$ is
\begin{equation}
    \Var[N(t)] = \frac{t}{\mu^3}(\sigss^2 + 2 \gammass) = \frac{t}{\mu^3}(\sigss^2 + 2 \sum_{l=1}^{\infty} \gamma_l).
    \label{suppeq:count_var_rewrite}
\end{equation}
The convergence of covariance terms $\{\gamma_l\}$ varies significantly across signal–background–dead-time combinations $(S,B,t_d)$, as shown in \cref{suppfig:convergence_dynamics}. Even with fixed $(S,B)$, changing $t_d$ can drastically alter both the rate and the oscillation pattern of convergence.  Consequently, it is nontrivial to select a single cutoff $c$ that generalizes to all operating conditions.  A larger $c$ improves accuracy but incurs heavy computation, since evaluating $\mPtilde^{\,l-1}$ in \cref{eq:covar_ss_supp} involves repeated large matrix exponentiations. Smaller $c$ reduces runtime but induces bias in long-range correlations.

A naive idea is to adopt a data-adaptive stopping rule: computing covariance terms until $\gamma_l$ falls below a threshold. However, this approach fails for two reasons: (1) $\gamma_l$ often exhibits oscillatory decay, repeatedly crossing zero, so early stopping can discard essential long-tail contributions; and (2) these missed tails are further amplified when the prefactor $t/\mu^3$ is large, such as in low dead-time regimes.

\begin{figure*}[t]
    \centering
    \includegraphics[width=\linewidth]{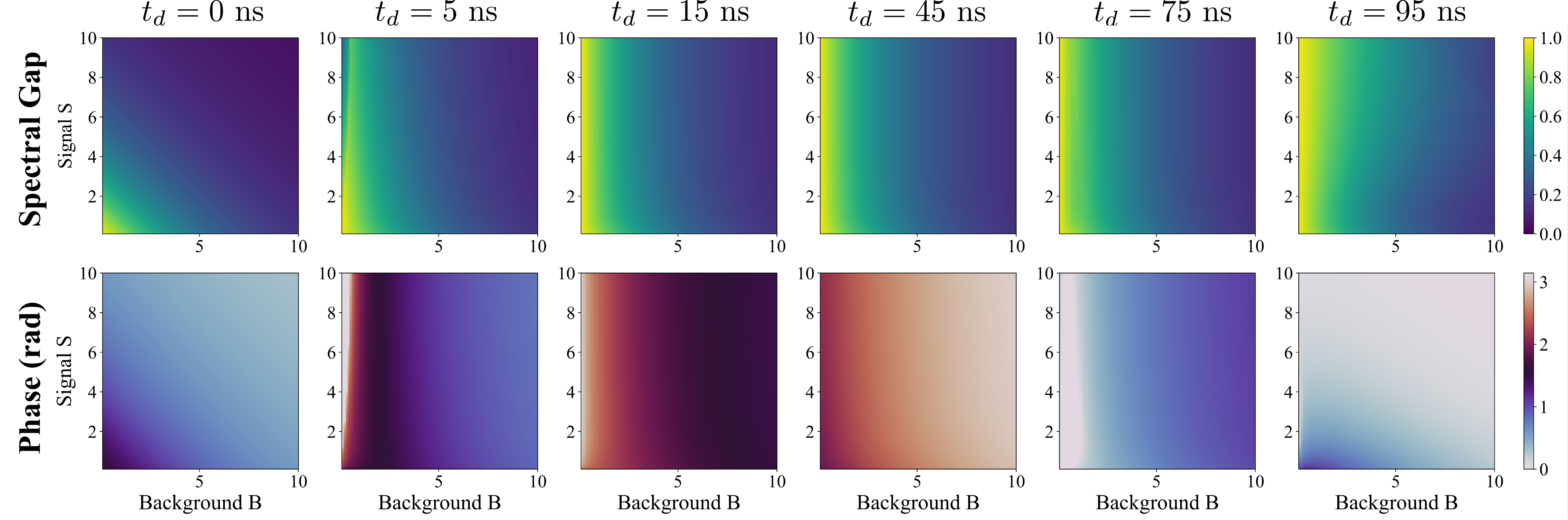}
    \caption{
    \textbf{Spectral Gap and Phase of the Second-Largest Eigenvalue $\lambda_2$ of $\mPtilde$ under Different $(S,B,t_d)$.}
    The spectral gap is defined as $1 - |\lambda_2|$, where $|\lambda_2|$ is the magnitude of the eigenvalue; a larger gap indicates faster decay of long-lag covariances.  
    The phase $\angle\lambda_2$ controls oscillatory behavior in the covariance sequence, with larger phases leading to more rapid oscillations.  
    Both quantities vary irregularly across operating conditions, making it difficult to design a universal covariance cutoff rule.
    }
    \label{suppfig:covariance_truncation}
\end{figure*}

Zhang \etal~\cite{zhang_2025_PDFacceleration} showed that the convergence rate and oscillations of Markov-chain covariances are dictated by the spectral properties of $\mPtilde$—in particular, the magnitude and phase of its second-largest eigenvalue $\lambda_2$. As shown in \cref{suppfig:covariance_truncation}, the spectral gap and phase vary irregularly across different $(S,B,t_d)$ settings, making it difficult to derive a universal deterministic cutoff or a reliable adaptive stopping rule.

\subsection{Low-Rank Spectral Approximation}
If $\mPtilde$ were diagonalizable, we could write
\begin{equation}
    \mPtilde = \mV \mLambda \mUH, 
    \quad 
    \mPtilde^{\,l-1} = \mV \mLambda^{\,l-1} \mUH,
    \label{suppeq:eigendecomp_P}
\end{equation}
where $\mLambda = \mathrm{diag}(\lambda_1, \lambda_2, \ldots)$ stores all eigenvalues and $\mV$ and $\mU$ contain the right and left eigenvectors, respectively, 
and satisfy $\mUH\mV = \mathbf{I}$.

Under this assumption, the steady-state covariance $\gammass$ in \cref{eq:covar_ss_supp} becomes
\begin{equation}
    \gammass = \sum_{m=2}^M \frac{\alpha_m \beta_m}{1 - \lambda_m},
    \label{suppeq:gammass_eigendecomp}
\end{equation}
where $M$ is the rank of $\mPtilde$, $\alpha_m \coloneqq (\vpi \mQ_{\vmu}\mV)_m$, and $\beta_m \coloneqq (\mUH\mQ_{\vmu}\mathbf{1})_m$.
\medskip

\noindent \textbf{Proof of \cref{suppeq:gammass_eigendecomp}.}
Plugging $\mPtilde^{\,l-1}$ in \cref{suppeq:eigendecomp_P} into the expression of $\gamma_l$ in \cref{suppeq:cov_l_expression},
\begin{align}
    \gamma_l
    & = \underset{\valpha^\top}{\underbrace{\vpi \mathbf{Q_{\vmu}} \mV}} \mLambda^{\,l-1} \underset{\vbeta}{\underbrace{\mUH \mathbf{Q_{\vmu}} \mathbf{1}}} - \mu^2 \notag \\
    & = \sum_{m=1}^M \lambda_m^{l-1} \alpha_m \beta_m - \mu^2 \notag \\
    & \overset{(i)}{=} \sum_{m=2}^M \lambda_m^{l-1} \alpha_m \beta_m,
    \label{suppeq:gamma_l_spectral}
\end{align}
where $(i)$ is based on
\[
    \lambda_1^{l-1} \alpha_1 \beta_1 = 1 \cdot \vpi \mathbf{Q_{\vmu}} \mathbf{1} = \mu^2.
\]
Substituting \cref{suppeq:gamma_l_spectral} into the definition of $\gammass$ in \cref{eq:covar_ss_supp},
\begin{align}
    \gammass
    & = \sum_{l=1}^{\infty} \gamma_l = \sum_{l=1}^{\infty} \sum_{m=2}^M \lambda_m^{l-1} \alpha_m \beta_m \notag \\
    & = \sum_{m=2}^M \left( \sum_{l=1}^{\infty} \lambda_m^{l-1}\right) \alpha_m \beta_m \notag \\
    & \overset{(i)}{=} \sum_{m=2}^M \frac{\alpha_m \beta_m}{1 - \lambda_m},
    \label{suppeq:gammass_eigen_spectral}
\end{align}
where $(i)$ holds because all $\lambda_m$ are strictly smaller than $1$ and the equation of infinite geometric series is applied.
\hfill $\square$
\medskip

In practice, however, $\mPtilde$ is large and typically defective, lacking a complete eigenbasis. 
Computing its Jordan form or exact powers is infeasible. 
We therefore approximate $\mPtilde$ using a low-rank spectral decomposition.

We compute the leading $p$ eigenpairs $\{(\lambda_i, \vv_i, \vu_i)\}_{i=1}^{p}$ of $\mPtilde$, enforcing biorthogonality $\mUH_p\mV_p = \mathbf{I}_p$:
\[
    \mPtilde \approx \mV_p \mLambda_p \mUH_p, 
    \quad
    \mPtilde^{\,l-1} \approx \mV_p \mLambda_p^{\,l-1} \mUH_p.
\]
As $p$ is supposed to be small, this low-rank decomposition will be more accurate to model higher-order covariance terms, where inferior eigenvalues already converge to $0$. This inspires us to prescribe a cutoff $L$, calculate the first $L$ covariance terms as what they are, and include the low-rank decomposition to account for long-tail behavior, as stated in Proposition~\textcolor{cvprblue}{1}:
\begin{equation}
    \widehat{\gammass}(L,p)
    =
    \underbrace{\sum_{l=1}^{L}\gamma_l}_{\text{exact low-order terms}}
    +
    \underbrace{\sum_{m=2}^p
    \frac{\alpha_m \beta_m \lambda_m^{L}}{1 - \lambda_m}}_{\text{spectral tail estimation}},
    \label{suppeq:gammass_spectral}
\end{equation}

\noindent \textbf{Proof of \cref{suppeq:gammass_spectral}.}
\begin{align}
    \widehat{\gammass}(L,p)
    & = \sum_{l=1}^{\infty} \gamma_l = \sum_{l=1}^{L}\gamma_l + \sum_{l=L+1}^{\infty}\gamma_l \notag \\
    & \overset{(i)}{=} \sum_{l=1}^{L}\gamma_l + \sum_{m=2}^p
    \frac{\alpha_m \beta_m \lambda_m^{L}}{1 - \lambda_m}, \notag
\end{align}
where $(i)$ follows a similar analysis of an infinite geometric series as in \cref{suppeq:gammass_eigen_spectral}.
\hfill $\square$

\subsection{Practical Choice of \texorpdfstring{$L$ and $p$}{L and p}}
\begin{figure}[t]
    \centering
    \includegraphics[width=\linewidth]{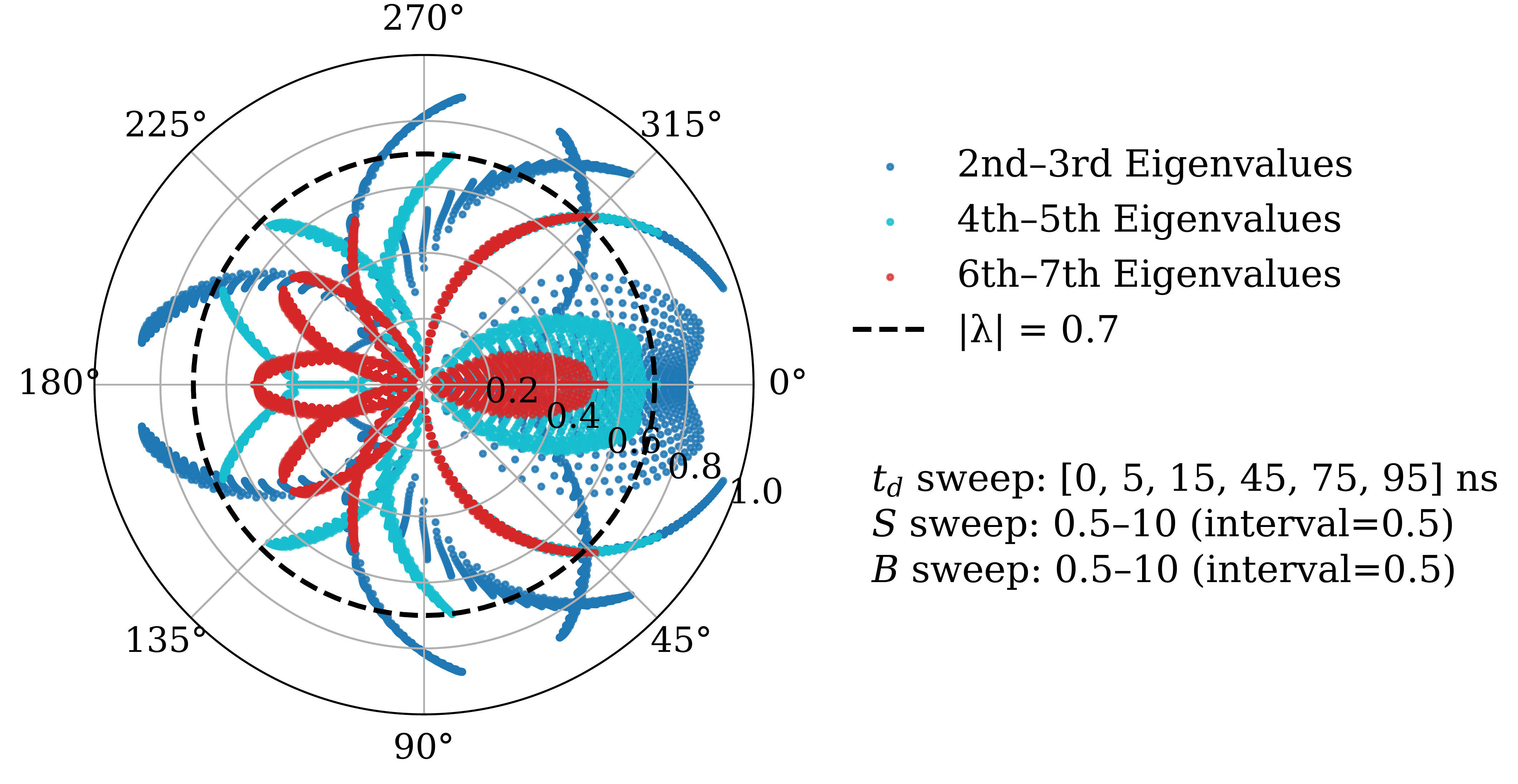}
    \caption{\textbf{Polar Plots of Leading Eigenvalues on an $(S,B,t_d)$ Grid.} This guides us to select $L = 6$ and $p = 5$ empirically.}
    \label{suppfig:polar}
\end{figure}

From \cref{suppeq:gammass_spectral}, the problem of estimating the steady-state covariance $\gammass$ accurately and efficiently boils down to choosing $L$ and $p$ wisely. We compute a few leading eigenvalues under common $(S,B,t_d)$ combinations, as illustrated in \cref{suppfig:polar}. Besides the largest eigenvalue, which is always $1$, they are usually conjugate, appearing in pairs. The largest magnitude of the $6$th and $7$th eigenvalues is around $0.7$. Their effects of the $6$th and $7$th and all remaining smaller eigenvalues will be negligible after six steps, since $0.7^6 \approx 0.1$. Therefore, we select $L = 6$ and $p = 5$, i.e. we calculate the first six covariance terms and estimate the remaining tail by four leading eigenmodes.

\section{Count Statistics v.s. Bin Resolution}
The transition matrix is the core component of our MRP framework, and its accuracy depends
on the discretization resolution of the registration timeline. A finer bin resolution yields a more faithful approximation of the continuous-time dynamics, but increases computational and memory cost. In this section, we study how the number of bins affects the accuracy and efficiency of the predicted count statistics.

We simulate an illumination cycle with $N_r = 1000$, dead time $t_d = 75$\,ns, signal level
$S = 8.2$, and background level $B = 1.2$. The transition matrix is constructed using
five different bin resolutions, from $2^{10}$ to $2^{14}$ states. The goal is to evaluate how
the predicted mean, variance, and runtime change as the discretization becomes finer.

\begin{figure*}[t]
    \centering
    \includegraphics[width=\linewidth]{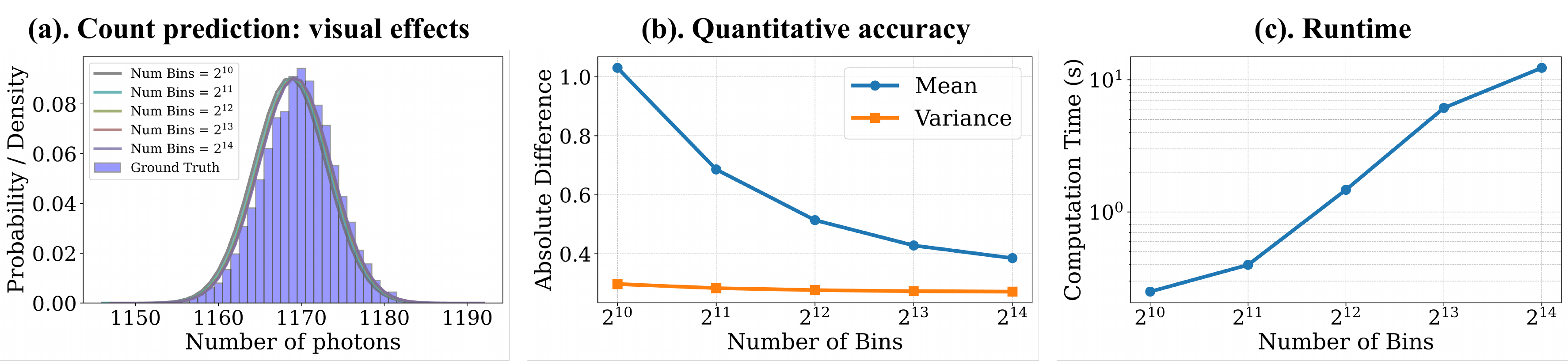}
    \caption{
    \textbf{Effect of Bin Resolution on Count Statistics Prediction.}
    }
    \label{suppfig:bin_resolution}
\end{figure*}

\medskip
\noindent\textbf{Accuracy.}
Figure~\ref{suppfig:bin_resolution}(a) shows the predicted count distributions. As the number of
bins increases, the predicted curves move consistently closer to the Monte Carlo ground
truth, generated by the gold-standard simulator, illustrating the improved fidelity of the discrete approximation. Figure~\ref{suppfig:bin_resolution}(b) quantifies this behavior: both the absolute mean error and variance error decrease
monotonically as bin resolution grows. For this specific SBR value, the mean prediction error
is relatively small, but this is not always the case under other SBRs. Moreover, since the mean
scales linearly with $N_r$, the impact of a fixed per-cycle error becomes more pronounced at
larger illumination counts.

\medskip
\noindent\textbf{Efficiency.}
Figure~\ref{suppfig:bin_resolution}(c) reports the runtime for constructing the transition matrix
and computing the statistics. As expected, higher bin resolutions incur additional cost due
to the larger matrix size. In practice, this cost is often acceptable because the transition
matrix and lookup tables can be precomputed once and reused across frames or scenes. For
applications requiring on-the-fly computation, a moderate bin resolution may offer a good
tradeoff between accuracy and speed.

All experiments in this section are performed on a CPU. Large-matrix operations in the MRP
pipeline are highly parallelizable, and substantial acceleration is expected on GPUs, which we leave for future work.

\section{Reflectivity Estimation}
\textbf{Dataset.}
For downstream reflectivity evaluation, we use the NYU Depth V2 dataset to simulate histogram cubes from paired RGB and depth images~\cite{Silberman_ECCV12_dataset}. The RGB intensity is normalized to serve as ground-truth reflectivity, and noisy photon histograms are generated using our forward model. We curate a diverse subset of scenes to cover a wide range of textures and depths, selecting $324$ different types of scenes in total, with $280$ for training, $30$ for validation, and $15$ for testing. This setup provides a controlled yet challenging benchmark for reflectivity reconstruction.

\begin{figure*}[t]
    \centering
    \includegraphics[width=\linewidth]{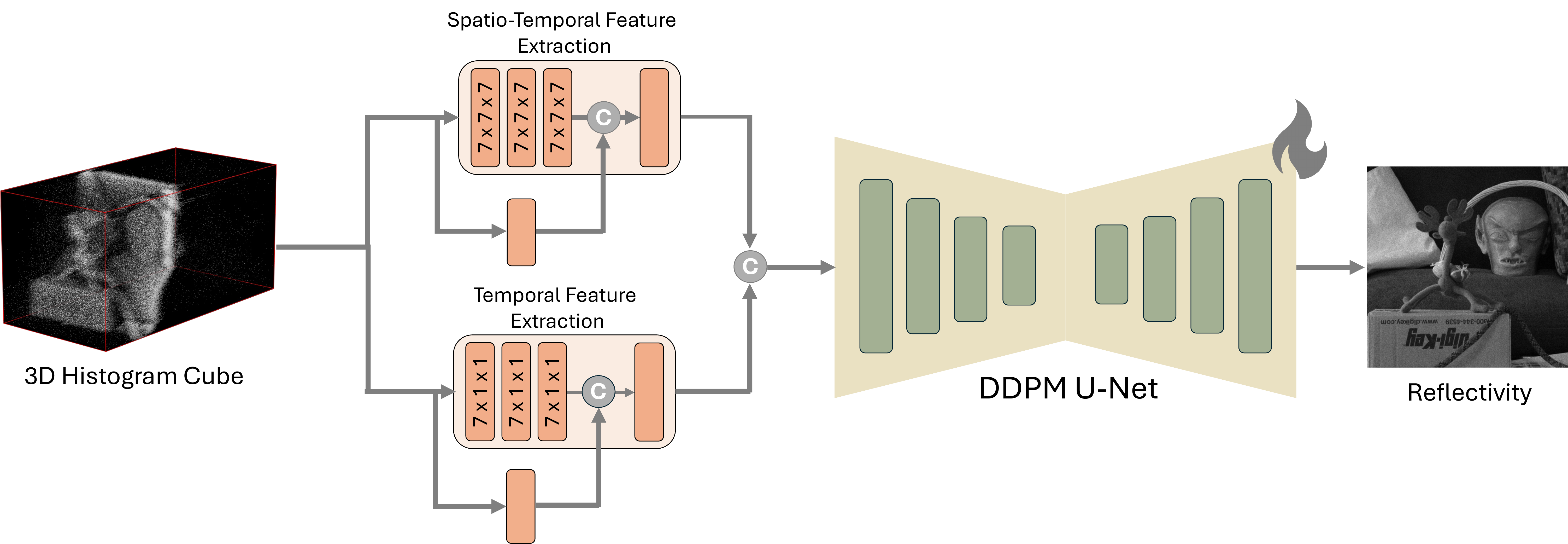}
    \caption{\textbf{Architecture utilized for Reflectivity Estimation.} The image used for illustration is from the Middlebury Stereo Datasets~\cite{Scharstein_2007_Middlebury}.} 
    \label{suppfig:architecture}
\end{figure*}

\noindent \textbf{Network Details.} To empirically analyze the accuracy of the underlying photon count estimator, we look into the task of reflectivity estimation as it is heavily dependent on the photon counts. As shown in \cref{suppfig:architecture}, the architecture utilized for the estimation is built upon the existing DDPM~\cite{ddpm}'s U-Net. For the model to effectively handle the $3$D Histogram Cube, we adapt the existing U-Net by introducing parallel residual modules. To effectively extract useful information from the input data, we employ a dual-branch feature extraction mechanism. 

The input is processed simultaneously by a \textbf{Spatio-Temporal Feature Extraction} module using $7 \times 7 \times 7$ kernels and a \textbf{Temporal Feature Extraction} module utilizing $7 \times 1 \times 1$ kernels. The latter branch is specifically designed to disentangle temporal features from spatial context; by operating solely along the temporal axis, the $7 \times 1 \times 1$ convolutions capture the precise time-evolution of the signal for each pixel independently, avoiding the smoothing of sharp temporal dynamics that typically occurs when aggregating spatial neighbors in standard $3$D convolutions. The features from these complementary branches are subsequently concatenated and fed into the \textbf{DDPM U-Net} to generate the final reflectivity map.

\noindent \textbf{Training Details.} We conduct separate training sessions for each of the four simulation environments: Poisson, Renewal, Zhang et al.~\cite{zhang_2025_icip}, and our proposed MaRS framework. The models are optimized using the Adam optimizer with a learning rate of $1 \times 10^{-4}$ and a weight decay of $1 \times 10^{-4}$ to ensure stable convergence. The training process is executed on a single NVIDIA A100 GPU with a batch size of $2$, processing input patches with a spatial resolution of $128 \times 128$ and a temporal depth of $1024$ bins. Each model is trained for $30,000$ epochs, requiring approximately two days of computation time. To preserve both radiometric accuracy and high-level structural fidelity, the network is supervised using a composite objective function defined as $\mathcal{L} = \lambda_{1}\mathcal{L}_{1} + \lambda_{2}\mathcal{L}_{\text{VGG}}$, where the pixel-wise $L_1$ loss and the VGG-based perceptual loss are assigned weights of $\lambda_{1}=1$ and $\lambda_{2}=0.2$, respectively.

\begin{figure}[t]
    \centering
    \includegraphics[width=\linewidth]{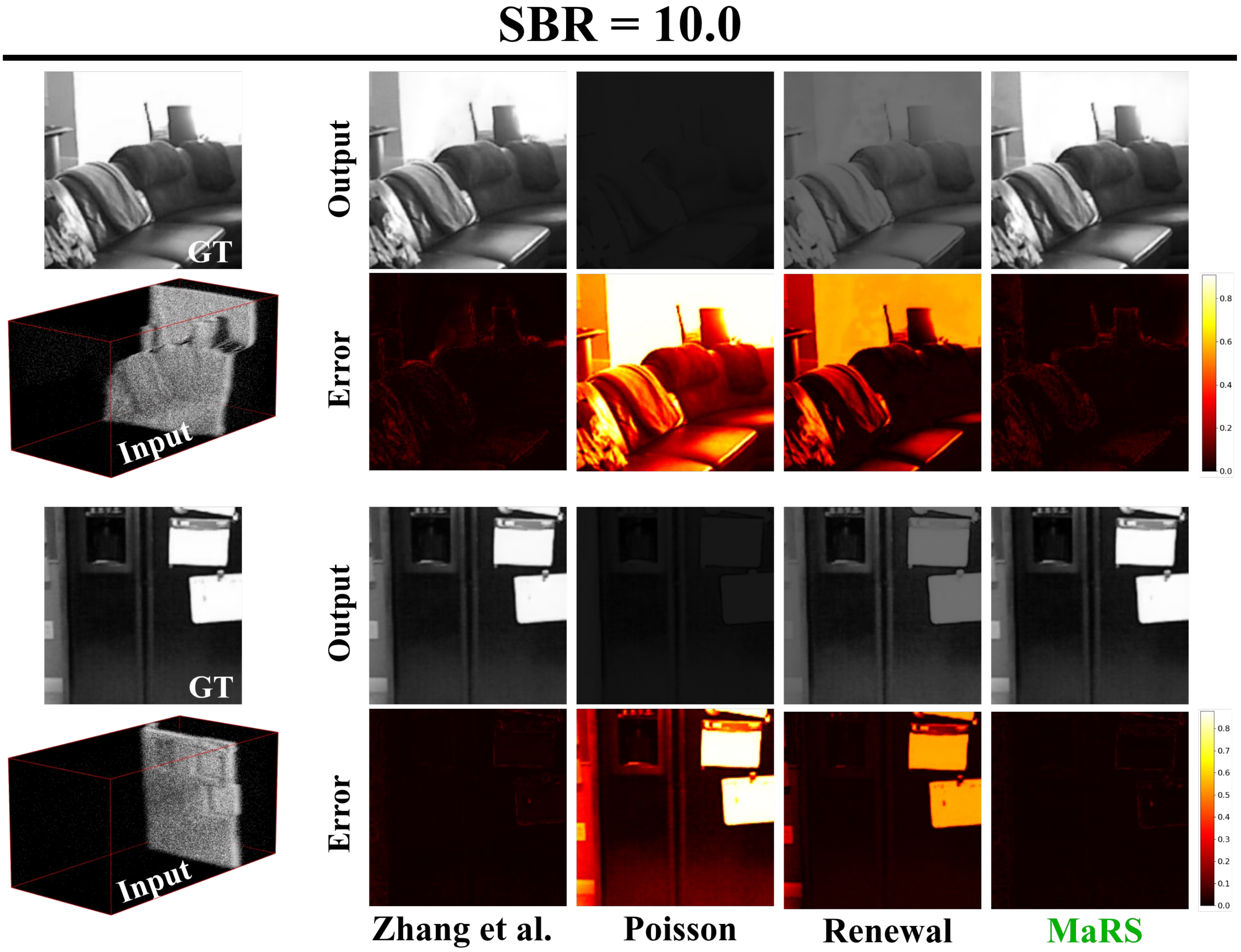}
    \caption{\textbf{Reflectivity Estimation Results for SBR = $10.0$.}}
    \label{suppfig:visualresults}
\end{figure}

\noindent \textbf{Testing Details.} During the inference phase, we evaluate the generalization capability of our models using a test set comprising $15$ diverse scenes derived from the NYU Depth V2  dataset~\cite{Silberman_ECCV12_dataset}. To ensure a comprehensive assessment, each model is evaluated under two distinct conditions: first, within its native training environment (the same simulator used for optimization), and second, against a ``Gold Standard'' simulator designed to rigorously replicate real-world acquisition physics. The testing input dimensions are set to $200 \times 200 \times 1024$. Quantitative performance is measured using Peak Signal-to-Noise Ratio (PSNR) and Structural Similarity Index Measure (SSIM) to evaluate pixel-level signal fidelity, alongside the Learned Perceptual Image Patch Similarity (LPIPS) metric to assess the perceptual quality of the reconstructed reflectivity maps.

\noindent \textbf{More Visual Results.} Detailed reflectivity estimation Visual Results for SBR $= 10.0$ has been provided in \cref{suppfig:visualresults}. We observe that despite an increase in Signal-to-Background Ratio (SBR), models trained on existing simulators output inferior outputs due to inaccurate modelling of photon counts, whereas model trained on our proposed simulator, MaRS, consistently outputs superior results.


\end{document}